\documentclass[preprint,showpacs,preprintnumbers]{revtex4}
\usepackage{graphicx}
\usepackage{dcolumn}
\usepackage{bm}

\begin{document}

\title{The LISA Time-Delay Interferometry Zero-Signal Solution. I:
  Geometrical Properties}

\author{Massimo Tinto} \email{Massimo.Tinto@jpl.nasa.gov}
\affiliation{Jet Propulsion Laboratory, California
  Institute of Technology, Pasadena, CA 91109}
\altaffiliation [Also at: ]{Space Radiation Laboratory, California
  Institute of Technology, Pasadena, CA 91125}

\author{Shane L. Larson}
\email{shane@srl.caltech.edu}
\affiliation{Space Radiation Laboratory, California Institute of Technology,
Pasadena, CA 91125}

\date{\today}

\begin{abstract}
  Time-Delay Interferometry (TDI) is the data processing technique
  needed for generating interferometric combinations of data measured
  by the multiple Doppler readouts available onboard the three LISA
  spacecraft. Within the space of all possible interferometric
  combinations TDI can generate, we have derived a specific
  combination that has zero-response to the gravitational wave signal,
  and called it the {\it Zero-Signal Solution} (ZSS). This is a
  two-parameter family of linear combinations of the generators of the
  TDI space, and its response to a gravitational wave becomes null
  when these two parameters coincide with the values of the angles of
  the source location in the sky.  Remarkably, the ZSS does not rely
  on any assumptions about the gravitational waveform, and in fact it
  works for waveforms of any kind. Our approach is analogous to the
  data analysis method introduced by G\"ursel \& Tinto in the context
  of networks of Earth-based, wide-band, interferometric gravitational
  wave detectors observing in coincidence a gravitational wave burst.
  The ZSS should be regarded as an application of the G\"ursel \&
  Tinto method to the LISA data.
\end{abstract}
  

\maketitle

\section{Introduction}

LISA, the Laser Interferometer Space Antenna, is a deep-space mission
of three spacecraft flying in a triangular formation.  Jointly
proposed to the National Aeronautics and Space Administration (NASA)
and the European Space Agency (ESA), it aims to detect and study
gravitational radiation in the millihertz frequency band via laser
interferometry \cite{PPA98}.

Modeling each spacecraft as carrying lasers, beam splitters,
photo-detectors, and drag-free proof masses on each of two optical
benches, it has been shown \cite{AET}, \cite{ETA}, \cite{TEA},
\cite{DNV}, \cite{Shaddock}, \cite{CH}, \cite{STEA}, \cite{TEA04} that
the six measured time series of Doppler shifts of the one-way laser
beams between spacecraft pairs, and the six measured shifts between
adjacent optical benches on each spacecraft, can be combined, with
suitable time delays, to cancel the otherwise overwhelming frequency
fluctuations of the lasers ($\Delta \nu/\nu \simeq 10^{-13}/ \sqrt{\rm
  Hz}$), and the noise due to the mechanical vibrations of the optical
benches (which could be as large as $\Delta \nu/\nu \simeq 10^{-16}/
\sqrt{\rm Hz}$). The achievable strain sensitivity level $h \simeq
10^{-21} / \sqrt{\rm Hz}$ is set by the buffeting of the drag-free
proof masses inside each optical bench, and by the shot noise at the
photo-detectors.

In contrast to Earth-based, equal-arm interferometers for
gravitational wave detection, LISA will have multiple readouts. The
data they generate, when properly time shifted and linearly combined,
provide observables that are not only insensitive to laser frequency
fluctuations and optical bench motions, but at the same time show
different couplings to gravitational radiation and to the remaining
system noises. The technique for synthesizing all these
interferometric combinations has been called Time-Delay Interferometry
(TDI), and it has been shown that the functional space it generates
can be obtained by properly time-shifting and linearly combining the
four Sagnac generators ($\alpha, \beta, \gamma, \zeta$) \cite{AET},
\cite{ETA}, \cite{DNV}, \cite{TAE}. In particular, there exists a
combination of the four generators that, for a given waveform and
source location in the sky, achieves maximum signal-to-noise ratio
(SNR) \cite{Prince02}.

Here we further elaborate on the geometrical properties of the four
generators' responses. We find that there exists a two-parameter
family of combinations of ($\alpha, \beta, \gamma, \zeta$) that has
null response to an observed gravitational wave signal. In the case of
gravitational wave bursts, or sinusoidal signals observed over a time
scale short enough that the amplitude and frequency modulation induced
on the received signal by the LISA motion around the Sun are
negligible, this new TDI combination shows a sharp null when its two
parameters coincide with the values of the angles describing the
source location in the sky, or its mirror image with respect to the
LISA plane. This combination, which we call the {\it Zero-Signal
  Solution} (ZSS), does not rely on any assumptions about the
gravitational waveform, and in fact it works for gravitational
waveforms of any kind. Although its ability to identify these two
points in the sky degrades at lower frequencies, it should be regarded
as an improvement over the Symmetric Sagnac combination, $\zeta$,
\cite{TAE} since it could be used for discriminating gravitational wave
signals from spurious fluctuations generated by instrumental noise in
the entire LISA frequency band.

The ability of the ZSS to identify the source location in the sky
follows from the fact that the time spent by the gravitational
wavefront to propagate across the LISA constellation enters into the
gravitational wave responses of the four generators ($\alpha, \beta,
\gamma, \zeta$).  The two independent travel times spent by the wave
to propagate from one spacecraft to the other two define the
``six-pulse'' \cite{AET} structure in the responses of the four
generators.  Since the signal contribution to the ZSS combination
becomes null at the correct values of these two time-delays, and
the antenna pattern of the ZSS is symmetric with respect to the LISA
plane, it follows that in general there are two possible points in the
sky where the signal must have originated.  If the duration of the
signal is such that the effects of the motion of the LISA array around
the Sun can be disregarded (as in the case of bursts), then the
two-fold degeneracy cannot be removed.  However, if the duration of
the signal is such that the effects of the motion of LISA around the
solar system barycenter can be measured (as in the case of signals
from binary systems), then the source location can be uniquely
identified.  As LISA moves around the Sun, the point in the sky where
the source is located will remain fixed, while the other will move in
such a way to remain its mirror image with respect to the LISA plane.
By constructing the ZSS with different stretches of data, the unique
source location can be identified.

The ability of the ZSS to identify two possible points in the sky
where an observed gravitational wave signal must have come from will
be particularly useful when observing gravitational wave bursts, or
continuous signals that are strong enough to be observed by
integrating the data over time scales shorter than the time scale
during which the phase and amplitude modulations of the signal become
measurable.  In the latter case, the ZSS should be regarded as the
first step of a hierarchical procedure for measuring the parameters of
a continuous signal. The ZSS could be particularly useful when
observing, for instance, gravitational radiation emitted during the
inspiral of two super-massive blackholes, or one smaller object
into a super-massive hole. The ability of the ZSS to identify the
location of the binary system in the sky, without relying on the
knowledge of the signal waveform, might reduce significantly the
number of templates needed in order to study waveforms from such
astrophysical systems \cite{CT}.

This paper is organized as follows.  Section II provides a brief
overview of Time-Delay Interferometry, the data analysis technique
needed to remove the frequency fluctuations of the six lasers and
other noise sources affecting the LISA measurements.  After showing
that the entire space of interferometric combinations can be generated
by properly combining the ``six-pulse'' Sagnac combinations ($\alpha,
\beta, \gamma, \zeta$), in Section III we obtain the general
expression of the ZSS in the Fourier domain, and note that it does not
rely on any assumptions about the time-dependence of the observed
gravitational wave signal. The corresponding expression of the ZSS in
the time-domain is then given in Section IV, where we show that the
three data sets, $\alpha$, $\beta$, and $\gamma$, enter with
eighty-one different time delays.  Section V derives the expression of
the ZSS in the long-wavelength approximation (i.e. when the
characteristic wavelength of the wave is much larger than the typical
linear size of the LISA array). In Section VI we give a summary of our
work and emphasize that this ``part I'' article only addresses the
geometrical properties of the ZSS. A quantitative analysis covering
its ability to discriminate gravitational wave signals from random
noise fluctuations affecting the LISA data, and the accuracies it will
achieve in reconstructing the source location in the sky and the
wave's two independent amplitudes, will be presented in a companion
article.

\section{Summary of time-delay interferometry for LISA}

A brief overview of TDI, the technique needed to remove the frequency
fluctuations of the lasers from phase measurements obtained with an
unequal-arm interferometer, is presented in this section.  The reader
is referred to the articles \cite{TA}, \cite{AET}, \cite{ETA},
\cite{TEA}, \cite{DNV} for a derivation of TDI valid for the
non-rotating and rigid LISA configuration (the so called ``first
generation TDI'' (TDI1)), and to the following ones \cite{Shaddock},
\cite{CH}, \cite{STEA}, \cite{TEA04} for the modified expressions
accounting for the LISA spacecraft motions (the ``second generation''
TDI (TDI2)). The overall LISA geometry is shown in Figure 1.
\begin{figure}
\centering
\includegraphics[width=2.5in, angle=-90]{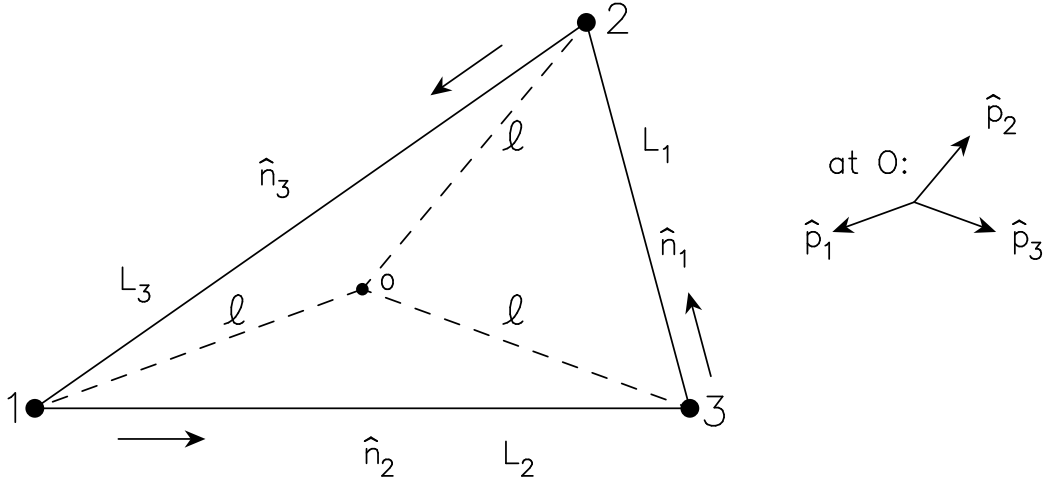}
\caption{Schematic LISA configuration. Each spacecraft is 
  equidistant from point o, with unit vectors $\hat p_i$ indicating
  directions to the three spacecraft. Unit vectors $\hat n_i$ point
  between spacecraft pairs with the indicated orientation.}
\end{figure}
There are six beams exchanged between the LISA spacecraft, 
with the six Doppler measurements $y_{ij}$ ($i,j = 1, 2, 3$) recorded
when each transmitted beam is mixed with the laser light of the
receiving optical bench.  The frequency fluctuations from the six
lasers, which enter in each of the six Doppler measurements, need to
be suppressed to a level smaller than that identified by the secondary
(proof mass and optical path) noises \cite{PPA98} in order to detect
and study gravitational radiation at the predicted amplitudes. 

Since the LISA triangular array has systematic motions, the two
one-way light times between any spacecraft pair are not the same
\cite{Shaddock}.  Delay times for travel between the spacecraft must
be accounted for depending on the sense of light propagation along
each link when combining these data as a consequence of the rotation
of the array.  Following \cite{TEA04}, the arms are labeled with
single numbers given by the opposite spacecraft; e.g., arm 2 (or
$2^{'}$) is opposite spacecraft 2, where primed delays are used to
distinguish light-times taken in the counter-clockwise sense and
unprimed delays for the clockwise light times (see Figure 2).  Also
the following labeling convention of the Doppler data will be used.
Explicitly: $y_{23}$ is the one-way Doppler shift measured at
spacecraft 3, coming from spacecraft 2, along arm 1.  Similarly,
$y_{32}$ is the Doppler shift measured on arrival at spacecraft 2
along arm $1'$ of a signal transmitted from spacecraft 3.  Due to the
relative motion between spacecraft, $L_1 \neq L_1^{'}$ in general. As
in \cite{ETA}, \cite{TEA}, we denote six further data streams,
$z_{ij}$ ($i,j = 1, 2, 3$), as the intra-spacecraft metrology data
used to monitor the motion of the two optical benches and the relative
phase fluctuations of the two lasers on each of the three spacecraft.
Finally, the light times $L_i$ and $L_i^{'}$ ($i = 1, 2, 3$) are not
only different (pure rotation) but also change in time. For this
reason, in the subscript notation for delays, their {\it{order}} is
important \cite{CH}, \cite{STEA}, \cite{TEA04}.
\begin{figure}
 \centering
\includegraphics[width=3.0 in, angle=0.0]{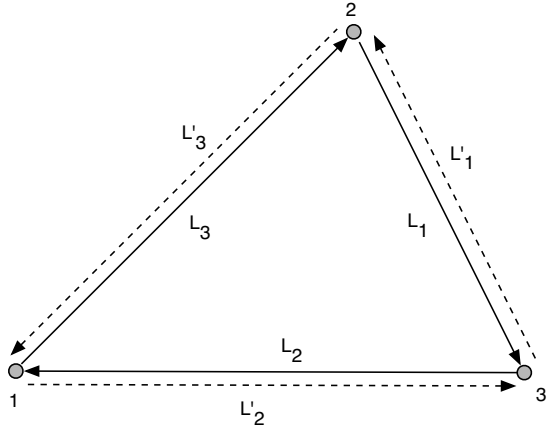}
\caption{Schematic diagram of LISA configurations involving
  six laser beams. Optical path delays taken in the counter-clockwise
  sense are denoted with a prime, while unprimed delays are in the
  clockwise sense.}
\end{figure}
The frequency fluctuations introduced by the gravitational wave
signal, the lasers, the optical benches, the proof masses, the fiber
optics, and the measurement itself at the photo-detector (shot noise)
enter into the Doppler observables $y_{ij}$, $z_{ij}$ with specific
time signatures. They have been derived in the literature \cite{ETA},
\cite{TEA}, and we refer the reader to those papers for a detailed
discussion.  The Doppler data $y_{ij}$, $z_{ij}$ are the fundamental
measurements needed to synthesize all the interferometric observables
unaffected by laser and optical bench noises.

Let us consider the TDI2 Sagnac observables: ($\rho_1, \rho_2,
\rho_3$)\footnote{In the literature \cite{STEA}, ($\rho_1, \rho_2,
  \rho_3$) are denoted ($\alpha_1, \alpha_2, \alpha_3$).  In order to
  avoid confusion with the notation introduced in appendix A, we will
  use the symbols ($\rho_1, \rho_2, \rho_3$) for indicating the TDI2
  Sagnac observables}.  Their expressions in terms of the Doppler
measurements $y_{ij}$, $z_{ij}$ are as follows \cite{STEA}
\begin{eqnarray}
\rho_1 & = & [y_{31} + y_{23;2} + y_{12;12} +
y_{21;312} + y_{32;3'312} + y_{13;1'3'312}] 
\nonumber \\
& & - [y_{21} + y_{32;3'} + y_{13;1'3'} +
y_{31;2'1'3'} + y_{23;22'1'3'} + y_{12;122'1'3'}]
\nonumber
\\
& & + {1 \over 2} [(z_{21} - z_{31}) - (z_{21} -
z_{31})_{;2'1'3'312}
+ (z_{32} - z_{12})_{;3'}
\nonumber \\
& & + (z_{32} - z_{12})_{;12} - (z_{32} - z_{12})_{;3'312}
- (z_{32} - z_{12})_{;122'1'3'}
\nonumber \\
& & + (z_{13} - z_{23})_{;2} + (z_{13} - z_{23})_{;1'3'}
- (z_{13} - z_{23})_{;22'1'3'}
\nonumber\\
& & - (z_{23} - z_{13})_{;1'3'312}] \ ,
\label{eq:1}
\end{eqnarray}
with $\rho_2$, $\rho_3$ following from Eq. (\ref{eq:1}) by
permutations of the spacecraft indices. The semicolon notation shown
in equation (\ref{eq:1}) emphasizes that the operation of sequentially
applying two or more delays to a given measurement is non-commutative,
and a specific order has to be adopted to adequately suppress the
laser noises \cite{STEA}, \cite{TEA04}. Specifically: $y_{ij;kl}
\equiv y_{ij} (t - L_l (t) - L_k (t - L_l)) \ne y_{ij;lk}$.

The expressions of the gravitational wave signal and the secondary
noise sources entering into $\rho_1$ will in general be different
from those entering into $\alpha$, the corresponding Sagnac observable
derived under the assumption of a stationary LISA array \cite{AET},
\cite{ETA}. However, the other remaining, secondary noises in LISA are
so much smaller, and the rotation and systematic velocities in LISA
are so intrinsically small, that index permutation may still be done
for them \cite{TEA04}. It is therefore easy to derive the following
relationship between the signal and secondary noises in $\rho_1$,
and those entering into the TDI1 combination $\alpha$ \cite{STEA},
\cite{TEA04}
\begin{equation}
\rho_1 (t) \simeq \alpha (t) - \alpha (t - L_1 - L_2 - L_3) \ ,
\label{eq:2}
\end{equation}
where $L_i \ \ , \ \ i=1, 2, 3$ are the unequal-arm lengths of the
stationary LISA array. Equation (\ref{eq:2}) implies that any data
analysis procedure and algorithm that will be implemented for the TDI2
combinations can actually be derived by considering the corresponding
TDI1 combinations. For this reason, from now on we will focus our
attention on the gravitational wave responses of the TDI1 observables
($\alpha, \beta, \gamma, \zeta$) \cite{AET}, \cite{ETA}.

\begin{figure}
 \centering
\includegraphics[width=5.0 in, angle=0.0]{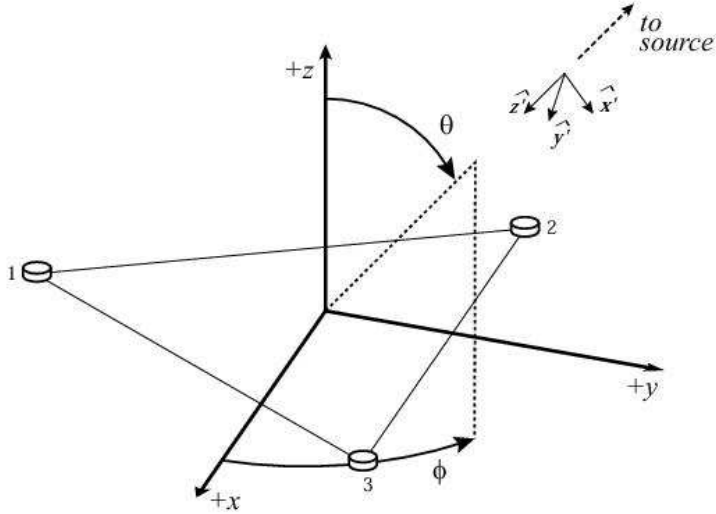}
\caption{The array and the wave coordinates. The $x$, $y$, and $z$
  axes are the LISA coordinate axes with the origin at the point
  $o$, and the array lies in the $x$, $y$ plane. The $x'$, $y'$, $z'$
  axes are the coordinate axes for the incoming gravitational
  wave. The $z'$ axis is parallel to the direction of propagation of the
  wave. The $x'$, $y'$ axes are those with respect to which $h_+
  (t)$ and $h_\times (t)$ are defined. The angles $\theta$, $\phi$
  describe the location of the source in the sky.}
\end{figure}
In the reference frame co-moving with LISA and centered at the point
$o$ that is equidistant from the three spacecraft (see Figure 3), the
response of $\alpha$ to a gravitational wave signal is given by the
following expression \cite{AET}
\begin{eqnarray}
\alpha^{gw}   & = &
\ \ {\left[ {1 - {\ l \over L_1} (\mu_2 - \mu_3)} \right]} \ (\Psi_1(t - \mu_2 \ l - 
L_1 - L_2) - \Psi_1(t - \mu_3 \ l - L_2))
\nonumber \\
& & -
{\left[ {1 + {\ l \over L_1} (\mu_2 - \mu_3)} \right]} \ (\Psi_1(t - \mu_3 \ l - 
L_1 - L_3) - \Psi_1(t - \mu_2 \ l - L_3))
\nonumber \\
& & +
\left[ 1 - {\ l \over L_2} (\mu_3 - \mu_1) \right] \ (\Psi_2(t - \mu_3 \ l - 
L_2) - \Psi_2(t - \mu_1 \ l))
\nonumber \\
& & -
{\left[ {1 + {\ l \over L_2} (\mu_3 - \mu_1)} \right]} \ (\Psi_2(t - \mu_1 \ l - 
L_2 - L_1 - L_3)
- \Psi_2(t - \mu_3 \ l - L_1 - L_3)) \ .
\nonumber \\
& & +
{\left[ {1 - {\ l \over L_3} (\mu_1 - \mu_2)} \right]} \ (\Psi_3(t - \mu_1 \ l - 
L_3 - L_1 - L_2)
- \Psi_3(t - \mu_2 \ l - L_1 - L_2))
\nonumber \\
& & -
{\left[ {1 + {\ l \over L_3} (\mu_1 - \mu_2)} \right]} \ (\Psi_3(t - \mu_2 \ l - 
L_3) - \Psi_3(t - \mu_1 \ l)) \ .
\label{eq:3}
\end{eqnarray}
Here $\mu_i = {\hat k} \cdot {\hat p_i }$, with $\hat p_i \ , \ i=1, 2, 3$
being the unit vector pointing from $o$ to spacecraft $i$, and $\hat k$
the unit propagation vector of the wave. The function $\Psi_i$
is equal to \cite{AET}
\begin{eqnarray} 
\Psi_i(t) & = & {1 \over 2} \ 
  \ {{{\hat n_i} \cdot {\bf h}(t) \cdot {\hat n_i}} \over { 1 - ({\hat
        k} \cdot {\hat n_i})^2}} \ ,
\label{eq:4}
\end{eqnarray} 
\noindent
where ${\bf h}(t)$ is the first order spatial metric perturbation at
point $o$.  Note that $L_1 {\hat k} \cdot {\hat n_1} = \ l (\mu_2 -
\mu_3)$, and so forth by cyclic permutation of the indices.  The
gravitational wave $\bf{h}$(t) can be written as $\left[ h_+(t) \ 
  {\bf{e_+}} + h_{\times}(t) \ {\bf{e_{\times}}} \right]$, where the
3-tensors $\bf{e_+}$ and $\bf{e_{\times}}$ are transverse to $\hat k$
and traceless.  With respect to an orthonormal propagation frame ($x',
y', z'$) their components are equal to
\begin{mathletters}
\begin{eqnarray}
\bf{e_+} & = &
\: \left(
\begin{array}{ccc}
1 & \; 0 & \; 0\\
0 & \; -1 & \; 0\\
0 & \; 0 & \; 0
\end{array} \; \;  \right)\;,
\;\bf{e_{\times}} = \left(
\begin{array}{ccc}
0 & \; 1 & \; 0\\
1 & \; 0 & \; 0\\
0 & \; 0 & \; 0
\end{array} \; \; \right) \ .
\label{eq:5}
\end{eqnarray}
\end{mathletters}
Equation (\ref{eq:3}) for $\alpha^{gw}$ shows a ``six-pulse response''
to gravitational radiation. In other words, a $\delta$-function
gravitational wave signal produces six distinct pulses in ($\alpha,
\beta, \gamma$) \cite{AET}, which are located with relative times
depending on the arrival direction of the wave and the detector
configuration.

Together with ($\alpha, \beta, \gamma$), the symmetric Sagnac
combination, $\zeta$, has been shown to form a basis for the TDI1
space of combinations and, like $\alpha$, has a six-pulse response to
gravitational radiation. Furthermore, it can be written in terms of
($\alpha, \beta, \gamma$) in the following way
\cite{AET}
\begin{equation}
\zeta - \zeta_{,123} = \alpha_{,1} - \alpha_{,23} + 
\beta_{,1} - \beta_{,23} + \gamma_{,1} - \gamma_{,23} \ .
\label{eq:6}
\end{equation}
The four interferometric combinations ($\alpha, \beta, \gamma, \zeta$)
jointly also give the expressions of the interferometric
combinations derived in \cite{AET}, \cite{ETA}: the Unequal-arm
Michelson (${\rm X}, {\rm Y}, {\rm Z}$), the Beacon (${\rm P}, {\rm
  Q}, {\rm R}$), the Monitor (${\rm E}, {\rm F}, {\rm G}$), and the
Relay (${\rm U}, {\rm V}, {\rm W}$) responses
\begin{eqnarray}
{\rm X}_{,1} & = & \alpha_{,23} - \beta_{,2} - \gamma_{,3} + \zeta \ ,
\label{eq:6a} \\
{\rm P} & = & \zeta - \alpha_{,1} \ ,
\label{eq:6b} \\
{\rm E} & = & \alpha - \zeta_{,1} \ ,
\label{eq:6c} \\
{\rm U} & = & \gamma_{,1} - \beta \ .
\label{eq:6d}
\end{eqnarray}
The remaining expressions can of course be obtained from equations
(\ref{eq:6a}, \ref{eq:6b}, \ref{eq:6c}, \ref{eq:6d}) by permutation of the
spacecraft indices. All these interferometric combinations have been
shown to add robustness to the mission with respect to failures of
subsystems, and potential design, implementation, or cost advantages
\cite{AET}, \cite{ETA}.

One important property of the $\zeta$ combination \cite{TAE} is that
it is much less sensitive to gravitational waves than any of the other
Sagnac combinations ($\alpha, \beta, \gamma$) in the low part of the
frequency band accessible to LISA ($f \alt 5$ mHz). Since, however, it
is affected by the same instrumental noise sources, it was recognized
that it can be used as a ``gravitational wave shield'' for estimating
the spectra of the instrumental noises affecting the ``sensitive''
combinations \cite{TAE}, \cite{ST}.

Although the $\zeta$ combination will be a very useful tool for
assessing the on-flight-noise performance of LISA \cite{TAE},
\cite{ST}, the question of whether there exists a TDI combination that
completely suppresses the gravitational wave signal still remains
open. In what follows we will tackle and solve this problem.

\section{The ZSS in the Fourier Domain}

LISA has the capability to {\underbar {simultaneously}} observe a
gravitational wave signal with several interferometric combinations of
different antenna patterns and noise transfer functions \cite{ETA}.
For this reason it should be regarded as an overlapping array of
interferometric gravitational wave detectors working in coincidence.
For a given signal waveform and source location in the sky, it has in
fact been shown \cite{Prince02} that there exists an {\it optimal}
combination of the generators of the TDI space that achieves maximum
SNR.

In what follows we will answer a complementary question to the one
addressed in \cite{Prince02}. We will identify, within the LISA TDI
functional space, a specific interferometric combination that has
zero-response (and therefore zero SNR) to a gravitational wave signal.
Such a combination provides a way for discriminating a gravitational
wave signal from noise-generated fluctuations, and for this reason it
should be regarded as an improvement over the Symmetric Sagnac
combination $\zeta$ \cite{TAE}.

Our approach is closely related to the one introduced by G\"ursel and
Tinto \cite{GT} in the context of data analysis for networks of
ground-based interferometers observing in coincidence a gravitational
wave burst.  There, a network of three ground-based, wide-band,
detectors widely separated on the Earth was shown to provide enough
information for uniquely solving the ``Inverse Problem'' in
gravitational wave astronomy: the reconstruction of the wave's two
independent amplitudes and the determination of the location of the
wave source in the sky. With three non-coplanar detector responses and
two independent time delays, it is possible to identify a unique,
two-parameter linear combination of the three independent detector
measurements that has zero-response to the gravitational wave signal
when the two parameters coincided with the values of the polar angles
of the source location in the sky \cite{GT}.

For LISA there exists an analogous two-parameter linear combination of
the four generators of the TDI space, ($\alpha, \beta, \gamma,
\zeta$), which in the high-frequency part of the LISA band is capable
of identifying two points in the sky where the observed gravitational
wave burst must have come from. The property of this combination,
which we have named the {\it Zero-Signal Solution}, follows from
the fact that the two independent travel times spent by the burst
propagating across the three spacecraft define the separation of the
``six-pulses'' in the responses of the four generators ($\alpha,
\beta, \gamma, \zeta$) \cite{AET}.  Each travel time defines a circle
on the sky where the signal must have come from, and their
intersection reduces the source location to two possible points.
However, due to the co-planarity of the antenna patterns of the four
generators, the LISA ZSS is unable to resolve the two-solution
ambiguity for bursts, as was possible with networks of three
Earth-based laser interferometers \cite{GT}.

In order to derive the expression of the ZSS, let us consider in the
Fourier domain an arbitrary element of the TDI space, which we will
refer to as ${\widetilde \eta}$
\begin{equation}
{\widetilde \eta} (f) \equiv a_1 (f, {\vec \lambda}) \ {\widetilde{\alpha}} (f) \ +
\ a_2 (f, {\vec \lambda}) \ {\widetilde{\beta}} (f) \ +
\ a_3 (f, {\vec \lambda}) \ {\widetilde{\gamma}} (f) \ +
\ a_4 (f, {\vec \lambda}) \ {\widetilde{\zeta}} (f)  \ ,
\label{eq:7a}
\end{equation}
where the $\{a_i (f, \vec \lambda)\}^4_{i=1}$ are arbitrary complex
functions of the Fourier frequency $f$, and of a vector $\vec \lambda$
containing parameters characterizing the gravitational wave signal
(source location in the sky, waveform parameters, etc.) and the noises
affecting the four responses (noise levels, their correlations, etc.).
For a given choice of the four functions $\{a_i \}^4_{i=1}$,
${\widetilde \eta}$
gives an element of the functional space of interferometric
combinations generated by ($\alpha, \beta, \gamma, \zeta$).  The
corresponding expression of the signal-to-noise ratio,
$SNR_{\eta}^2$, of the combination $\widetilde \eta$ is equal to
\begin{equation}
SNR_{\eta}^2 = 
\int_{f_{l}}^{f_u} 
{
{|a_1 \ {\widetilde \alpha_s} + a_2 \ {\widetilde \beta_s} + a_3 \
  {\widetilde \gamma_s} + a_4 {\widetilde \zeta_s} |^2} 
\over
{\langle|a_1 \ {\widetilde \alpha_n} + a_2 \ {\widetilde \beta_n} +
  a_3 \ {\widetilde \gamma_n} + a_4 {\widetilde \zeta_n} |^2 \rangle} } \ df \ .
\label{eq:7b}
\end{equation}
In equation (\ref{eq:7b}) the subscript $n$ refers to the noise part
of (${\widetilde {\alpha}}, {\widetilde {\beta}}, {\widetilde
  {\gamma}}, {\widetilde {\zeta}}$), the angle brackets represent
noise ensemble averages, and the interval of integration ($f_l, f_u$)
corresponds to the frequency band accessible by LISA.  As a
consequence of the relationship between $\zeta$ and $\alpha$, $\beta$,
and $\gamma$ (equation (\ref{eq:6})), it has been shown (see equation
(10) of reference \cite{Prince02}) that equation (\ref{eq:7b}) can be
rewritten in the following equivalent form
\begin{equation}
SNR_{\eta}^2 = 
\int_{f_{l}}^{f_u} 
{
{|a_1 \ {\widetilde \alpha^{gw}} + a_2 \ {\widetilde \beta^{gw}} + a_3 \
  {\widetilde \gamma^{gw}} |^2} 
\over
{\langle|a_1 \ {\widetilde \alpha_n} + a_2 \ {\widetilde \beta_n} +
  a_3 \ {\widetilde \gamma_n} |^2 \rangle} } \ df \ .
\label{eq:7}
\end{equation}
For a given choice of the three functions $\{a_i \}^3_{i=1}$,
$SNR_{\eta}^2$ gives the corresponding SNR of an element of the TDI
space, whose Fourier transform is equal to ${\widetilde \eta} \equiv a_1 \ 
{\widetilde \alpha} + a_2 \ {\widetilde \beta} + a_3 \ {\widetilde
  \gamma}$. Our goal is therefore to identify, for a given
gravitational wave signal, the three functions $\{a_i \}^3_{i=1}$ that
give $SNR_{\eta}^2 = 0$. Since $SNR_{\eta}^2$ is a positive definite
functional, it can be zero if and only if the numerator of its
integrand is identically null, i.e.
\begin{equation}
{\widetilde \eta}^{gw} (f) \equiv a_1 \ {\widetilde{\alpha}}^{gw} (f) \ +
\ a_2 \ {\widetilde{\beta}}^{gw} (f) \ +
\ a_3 \ {\widetilde{\gamma}}^{gw} (f) = 0\ .
\label{eq:8}
\end{equation}
The expressions for $\alpha^{gw}$, $\beta^{gw}$ and $\gamma^{gw}$ can
be rewritten in the following form (see equations (\ref{eq:3}),
(\ref{eq:4}), and appendix B)
\begin{eqnarray}
{\widetilde{\alpha}}^{gw} (f) & = & \alpha_+ (f, \theta_s, \phi_s ) \
{\widetilde{h}_+} (f) \ + \ \alpha_\times (f, \theta_s, \phi_s ) \
{\widetilde{h}_\times} (f) \ ,
\label{eq:9a}
\\
{\widetilde{\beta}}^{gw} (f) & = & \beta_+ (f, \theta_s, \phi_s ) \
{\widetilde{h}_+} (f) \ + \ \beta_\times (f, \theta_s, \phi_s ) \
{\widetilde{h}_\times} (f) \ ,
\label{eq:9b}
\\
{\widetilde{\gamma}}^{gw} (f) & = & \gamma_+ (f, \theta_s, \phi_s ) \
{\widetilde{h}_+} (f) \ + \ \gamma_\times (f, \theta_s, \phi_s ) \
{\widetilde{h}_\times} (f) \ ,
\label{eq:9c}
\end{eqnarray}
where ($\theta_s, \phi_s$) are the two angles describing the location
of the source in the sky with respect to a LISA coordinate frame (see
Figure 3), and the functions $\alpha_+$, $\alpha_\times$, $\beta_+$,
$\beta_\times$, $\gamma_+$, $\gamma_\times$ are 
pattern functions, derived in appendix B. From these considerations it
follows that ${\widetilde \eta}^{gw}$ can be rewritten in the following form
\begin{eqnarray}
{\widetilde \eta}^{gw}(f) & = & [a_1 \ \alpha_+ (f, \theta_s, \phi_s)
\ + \ a_2 \ \beta_+ (f, \theta_s, \phi_s) \ + \ 
a_3 \ \gamma_+ (f, \theta_s, \phi_s)] \ 
{\widetilde{h_+}} (f) 
\nonumber
\\
&  + &  
[a_1  \ \alpha_\times (f, \theta_s, \phi_s)
\ + \ a_2 \ \beta_\times (f, \theta_s, \phi_s) \ + \ 
a_3 \ \gamma_\times (f, \theta_s, \phi_s)] \ 
{\widetilde{h_\times}} (f) \ .
\label{eq:10}
\end{eqnarray}
which is obtained by substituting equations (\ref{eq:9a}, \ref{eq:9b},
\ref{eq:9c}) into equation (\ref{eq:8}).  The wave's two independent
amplitudes, $h_+$, $h_\times$, are referred to the wave axes ($x',
y'$) which, without loss of generality, can be assumed to be oriented
in such a way that the $x'$-axis lies parallel to the LISA plane (the
$x-y$ plane). In order to have ${\widetilde \eta}^{gw} = 0$ for any
arbitrary pairs of wave's amplitudes $h_+$, $h_\times$, equation
(\ref{eq:10}) implies that the following homogeneous linear system of
two equations in three unknowns must be satisfied
\begin{eqnarray}
a_1 \ \alpha_+ (f, \theta_s, \phi_s ) \ + \ a_2 \ \beta_+ (f,
\theta_s, \phi_s ) \ + \ 
a_3 \ \gamma_+ (f, \theta_s, \phi_s ) & = & 0
\label{eq:11a}
\\
a_1  \ \alpha_\times (f, \theta_s, \phi_s )
\ + \ a_2 \ \beta_\times (f, \theta_s, \phi_s ) \ + \ 
a_3 \ \gamma_\times (f, \theta_s, \phi_s ) & = & 0
\label{eq:11b}
\end{eqnarray}
Since the rank of the matrix associated with the linear system above is
in general equal to two, it is easy to derive the expressions for
$a_1$, $a_2$, $a_3$ that give zero-response to a gravitational wave signal
at the source location
\begin{eqnarray}
a_1 (f, \theta_s, \phi_s) & = & \beta_+ (f, \theta_s, \phi_s ) \
\gamma_\times (f, \theta_s, \phi_s ) 
\ - \ \beta_\times (f, \theta_s, \phi_s) \ \gamma_+ (f, \theta_s, \phi_s) \ ,
\label{eq:12a}
\\
a_2 (f, \theta_s, \phi_s) & = & \gamma_+ (f, \theta_s, \phi_s) \ \alpha_\times (f, \theta_s, \phi_s) 
\ - \ \gamma_\times (f, \theta_s, \phi_s) \ \alpha_+ (f, \theta_s, \phi_s) \ ,
\label{eq:12b}
\\
a_3 (f, \theta_s, \phi_s) & = & \alpha_+ (f, \theta_s, \phi_s) \
\beta_\times (f, \theta_s, \phi_s) \ - \ \alpha_\times (f, \theta_s,
\phi_s) \ \beta_+ (f, \theta_s, \phi_s) \ .
\label{eq:12c}
\end{eqnarray}
Equations (\ref{eq:12a}, \ref{eq:12b}, \ref{eq:12c}) imply that the
following linear combination of the Fourier transforms of the three
generators ($\alpha, \beta, \gamma$)
\begin{eqnarray}
{\widetilde \eta} & \equiv & [\beta_+ (f, \theta, \phi) \
\gamma_\times (f, \theta, \phi) \ - \ \beta_\times (f, \theta,
\phi) \ \gamma_+ (f, \theta, \phi)] \ {\widetilde{\alpha}} (f) 
\nonumber
\\
& + & [\gamma_+ (f, \theta, \phi) \
\alpha_\times (f, \theta, \phi) \ - \ \gamma_\times (f, \theta,
\phi) \ \alpha_+ (f, \theta, \phi)] \ {\widetilde{\beta}} (f) 
\nonumber
\\
& + & [\alpha_+ (f, \theta, \phi) \
\beta_\times (f, \theta, \phi) \ - \ \alpha_\times (f, \theta,
\phi) \ \beta_+ (f, \theta, \phi)] \ {\widetilde{\gamma}} (f) \ ,
\label{eq:13}
\end{eqnarray}
is a function of the two parameters ($\theta,\phi$), and it has null
response to the gravitational wave signal observed in ($\alpha, \beta,
\gamma$) when $(\theta, \phi) \to (\theta_s, \phi_s)$, regardless of
the particular waveform considered.

Note that the expression of ${\widetilde \eta}^{gw}$ given in equation
(\ref{eq:13}) changes sign under reflection with respect to the LISA
plane.  This is because the functions $\alpha_+$, $\beta_+$,
$\gamma_+$ are invariant under reflection with respect to the LISA
plane, while $\alpha_\times$, $\beta_\times$, $\gamma_\times$ change
sign under the same operation (see appendices A and B for the
derivation of these symmetry properties). This implies that the
combinations $\alpha_+ \beta_\times - \alpha_\times \beta_+$, $\beta_+
\gamma_\times - \beta_\times \gamma_+$, $\gamma_+ \alpha_\times -
\gamma_\times \alpha_+$, which define the ZSS ${\widetilde \eta}$,
also change sign under reflection with respect to the LISA plane. In
the case of a gravitational wave burst, or if we assume to observe a
periodic signal for time scales shorter than the time during which the
Doppler effect induced by the LISA motion around the Sun begins to be
observable, we conclude that the ZSS will not be able to distinguish
between the source location in the sky and the point that is its
mirror image with respect to the LISA plane.

In the specific case of a sinusoidal signal, for which the two
independent polarization components are characterized in the source
rest reference frame by a constant frequency $f_0$, it is interesting
to calculate the time scale during which such a periodic signal will
still appear in the LISA data as a sinusoid. In the LISA reference
frame, any of the TDI responses will measure a sinusoidal signal whose
frequency is Doppler modulated by the LISA motion around the Sun in
the following way
\begin{equation}
f (t) = f_0 [1 - \frac{{{\hat k} \cdot {\vec V} (t)}}{c}] \ ,
\label{eq:13a}
\end{equation}
where $\hat k$ is the unit vector of propagation of the wave, $\vec V$
is the velocity of the LISA guiding center $o$ relative to the solar
system barycenter, and $c$ is the speed of light \cite{CR},
\cite{KTV}. If we take the root-mean-square (r.m.s.) value of the
frequency change given by equation (\ref{eq:13a}), averaged over all
possible incoming directions of the GW signal, we obtain the following
resulting r.m.s.  value of the Doppler change of the signal frequency
in the LISA reference frame
\begin{equation}
\Delta f \equiv \langle (f (t) - f_0)^2 \rangle ^{1/2} = \pm f_0 \
\frac{V_0}{2 c} \ ,
\label{eq:13b}
\end{equation}
where $V_0$ is the velocity amplitude of the almost sinusoidal motion
of the array around the Sun, and the angle brackets $\langle \rangle$
denote an ensemble average over source directions. If we take
$V_0\simeq 30$ km/s, and assume $f_0 = 10^{-2}$ Hz, the corresponding
r.m.s. frequency shift is equal to $\Delta f = 5.0 \ \times \ 10^{-7}$
Hz. In other words, if we would integrate the data for about $20$
days, the data frequency resolution would be equal to or larger than
the corresponding r.m.s.  frequency shift induced by the Doppler
modulation. Since the frequency shift induced by the amplitude
modulation is smaller than the Doppler frequency modulation at this
frequency $f_0$, (\cite{CR},\cite{KTV}), it follows that over a period
of about $20$ days, the expression of the ZSS can be used for
identifying two points in the sky where an observed sinusoidal signal
might have arrived from.
\begin{figure}
 \centering
\includegraphics[width=3.7 in, angle=0.0]{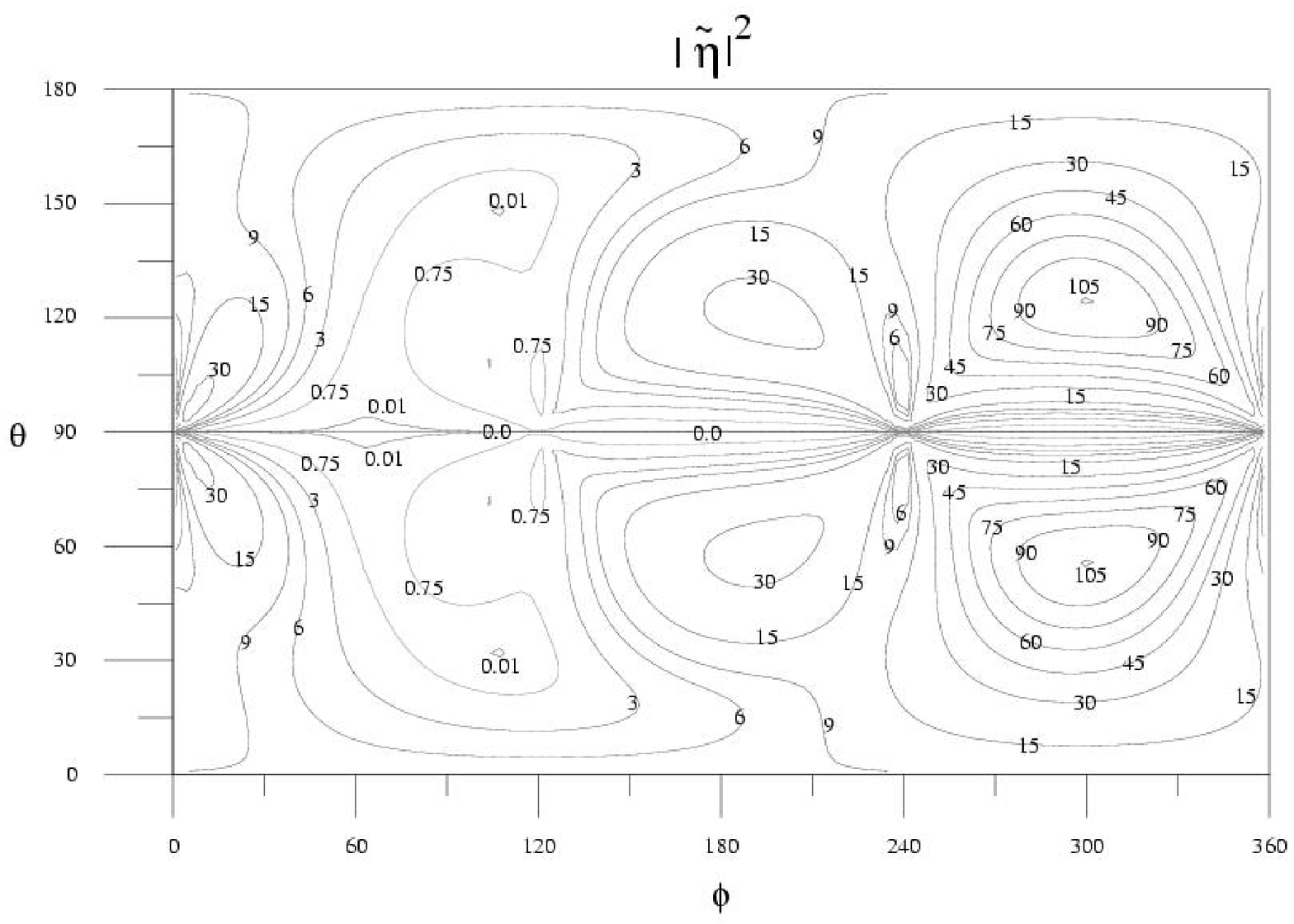}
\includegraphics[width=3.7 in, angle=0.0]{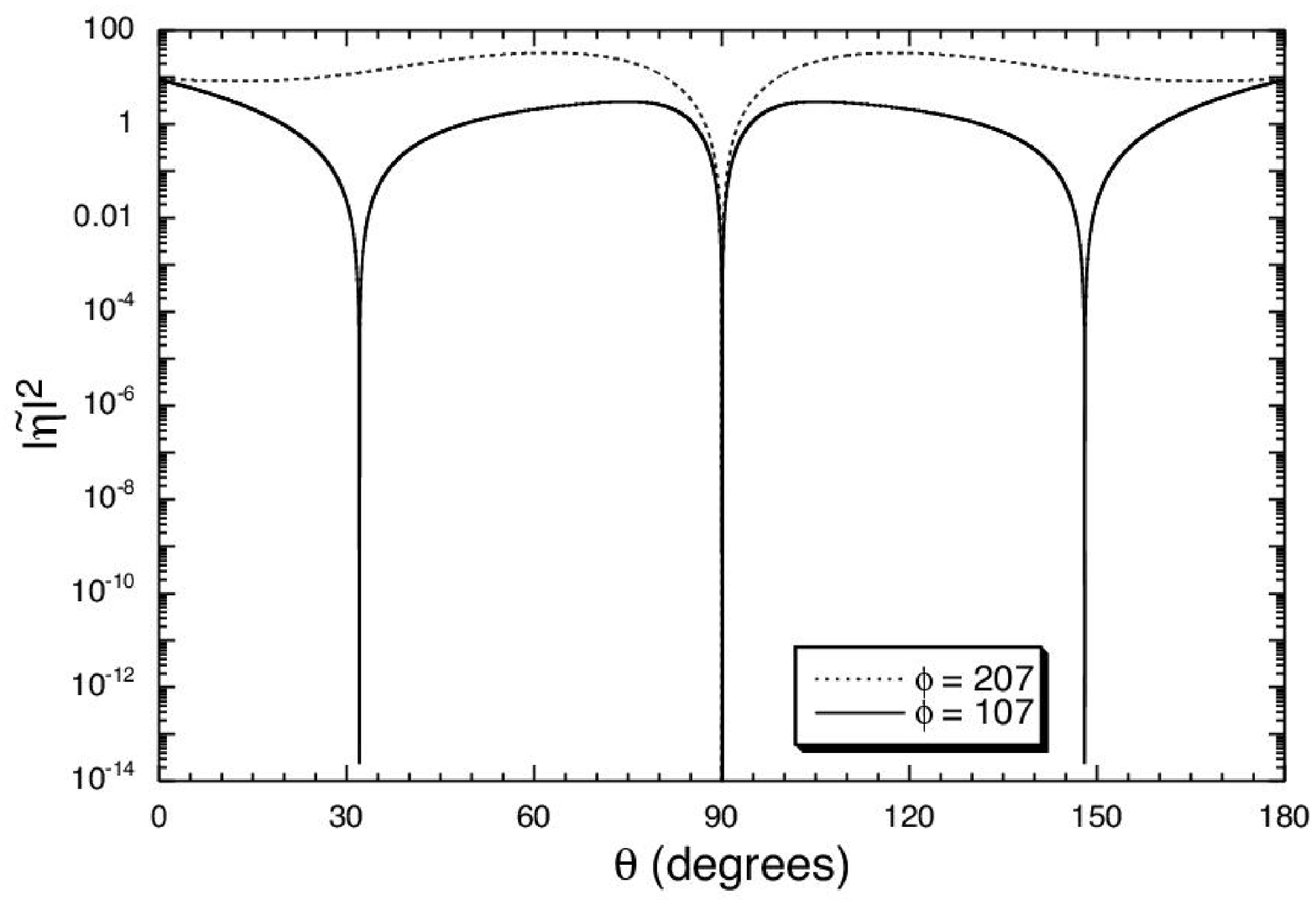}
\includegraphics[width=3.7 in, angle=0.0]{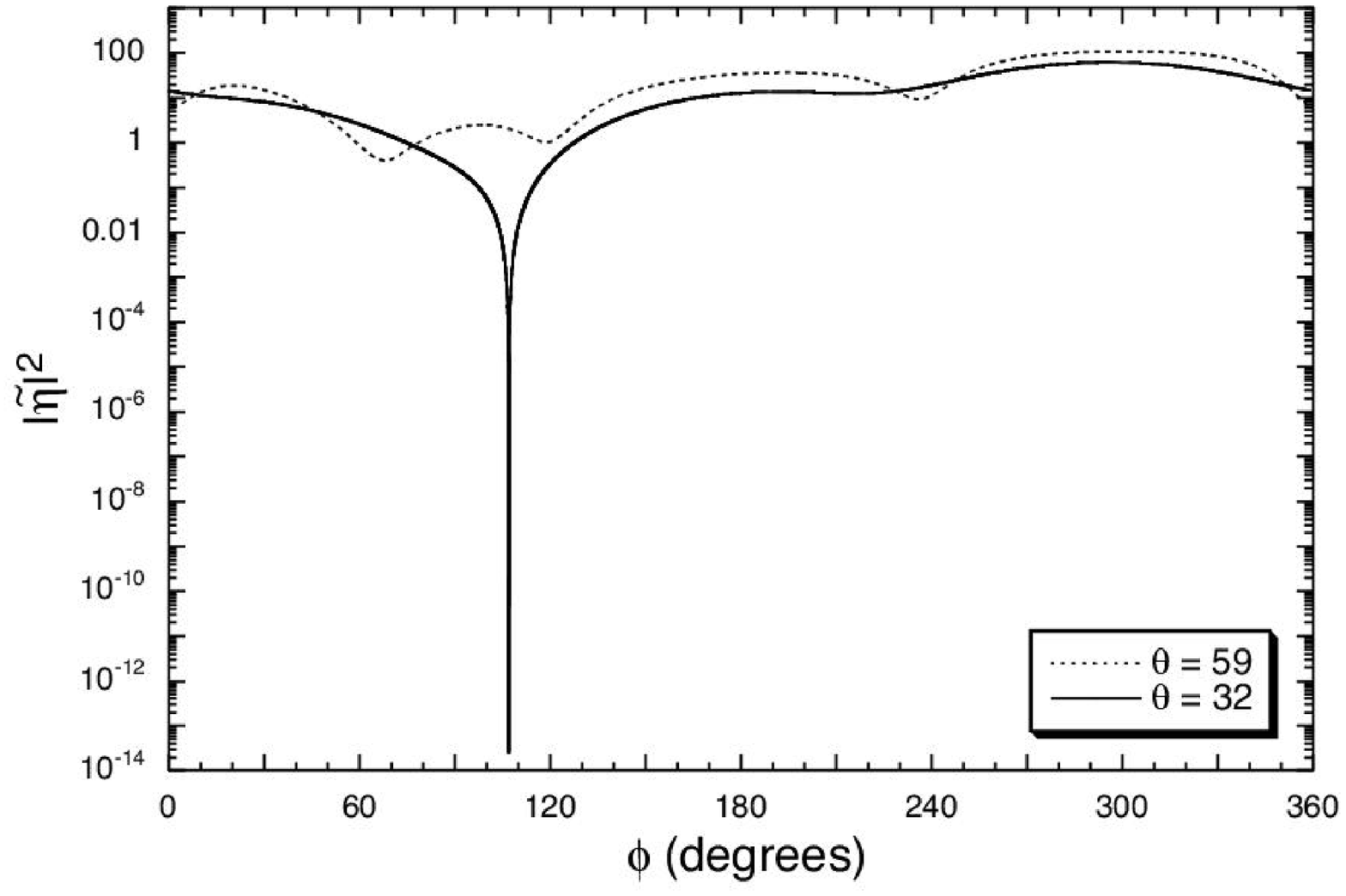}
\caption{(a) Contours of constant level of the modulus-squared of the ZSS
  constructed for a sinusoidal signal of frequency $f_0 = 10^{-2}$ Hz,
  incoming from a source located at: $\theta_s = 32^\circ \ , \ \phi_s
  = 107^\circ$. The plot is symmetric with respect to the LISA plane
  ($\theta = 90^\circ$); (b) two cross sections along the $\theta$-axis
  of the modulus-squared of the ZSS for the same signal of (a); one
  curve corresponds to $\phi = \phi_s = 107^\circ$, and the other
  to $\phi = 207^\circ \ne \phi_s$ (c) two cross sections along the
  $\phi$-axis, with $\theta = \theta_s = 32^\circ$, and $\theta =
  59^\circ$.}
\end{figure}

As an example, in Figure 4a we plot contours of constant power
spectral density of the ZSS in terms of its two-parameters, ($\theta,
\phi$), in the case of a circularly polarized sinusoidal signal of
frequency $f_0 = 10^{-2}$ Hz, incoming from $\theta_s = 32^\circ ,
\phi_s = 107^\circ$ and for noiseless data. Note that the ZSS is equal
to zero only at the correct source location, and at its mirror image
with respect to the LISA plane. Also, the plane of symmetry (defined
by $\theta = 90^\circ$) shows an identically null ZSS.  This is because
every TDI response is insensitive to the ``$\times$'' polarization
components of signals incoming from sources located in the LISA plane
(see Appendix A). This makes the ZSS degenerate for these source
directions.  Figure 4b shows instead the power spectral density of the
ZSS as a function of $\theta$, for the cases $\phi = \phi_s =
107^\circ$, and $\phi = 207^\circ$, while in Figure 4c we plot it as a
function of $\phi$ and for the choices $\theta = \theta_s = 32^\circ$,
and $\theta = 59^\circ$. From these two plots it is possible to have
an idea of the sharpness and narrowness of the nulls in terms of the
angles ($\theta, \phi$). In order to estimate the magnitude of the
accuracy by which these two points in the sky can be identified with
this method, however, it will be necessary to include the effects
induced by the noises and the magnitude of the SNR by which the signal
is detected in the three Sagnac observables ($\alpha, \beta, \gamma$)
\cite{GT}. This analysis will be presented in our forthcoming ``part
II'' article.
 
\section{The ZSS in the time domain}

The expression of the ZSS can also be given in the time-domain by
inverse Fourier transforming equation (\ref{eq:13}).  Having the ZSS in
the time domain will be particularly useful when attempting to
discriminate gravitational wave bursts against statistically
significant fluctuations generated by instrumental noise affecting the
three TDI combinations ($\alpha, \beta, \gamma$). A quantitative
analysis of the improvement in statistical confidence added by using
the ZSS as a veto against noise fluctuations will be addressed in a
follow up paper.  

In order to derive the time-domain expression of the ZSS, let us
consider only one of the three terms entering into the expression of
the Fourier transform of the ZSS given in equation (\ref{eq:13}), say
$a_3 (f, \theta, \phi) \ {\widetilde{\gamma}} (f)$.  This is because
the other two terms entering into the ZSS can be obtained from it by
permuting the spacecraft indices.

Since the function $a_3 (f, \theta, \phi)$ can be written in the
following analytic form
\begin{equation}
a_3 (f, \theta, \phi) = 
\sum_{k=1}^{27} A_3^{(k)} (\theta, \phi) \ e^{2 \pi i f \Delta_3^{(k)}} \ ,
\label{eq:14}
\end{equation}
where the twenty seven ``amplitudes'' $A_3^{(k)}$ and the
corresponding ``delays'' $\Delta_3^{(k)}$ depend only on the angles
($\theta, \phi$) and not on the Fourier frequency $f$, it follows from
the linearity of the inverse Fourier transform that the expression of
the ZSS in the time domain assumes the following form
\begin{equation}
\eta (t, \theta, \phi) = \sum_{k=1}^{27} 
[A_1^{(k)} \ \alpha(t - \Delta_1^{(k)}) + 
A_2^{(k)} \ \beta(t - \Delta_2^{(k)}) + 
A_3^{(k)} \ \gamma(t - \Delta_3^{(k)})] \ .
\label{eq:15}
\end{equation}
If we take into account that each of the Sagnac observable shows a
gravitational wave pulse at six-distinct times, we conclude that for
an arbitrary pair ($\theta, \phi$), the ZSS will display in principle
``four hundred eighty six'' pulses. At the correct source location,
and at its mirror image with respect to the LISA plane, the ZSS
becomes null, making all these pulses disappear.
\begin{figure}
 \centering
\includegraphics[width=3.7 in, angle=0.0]{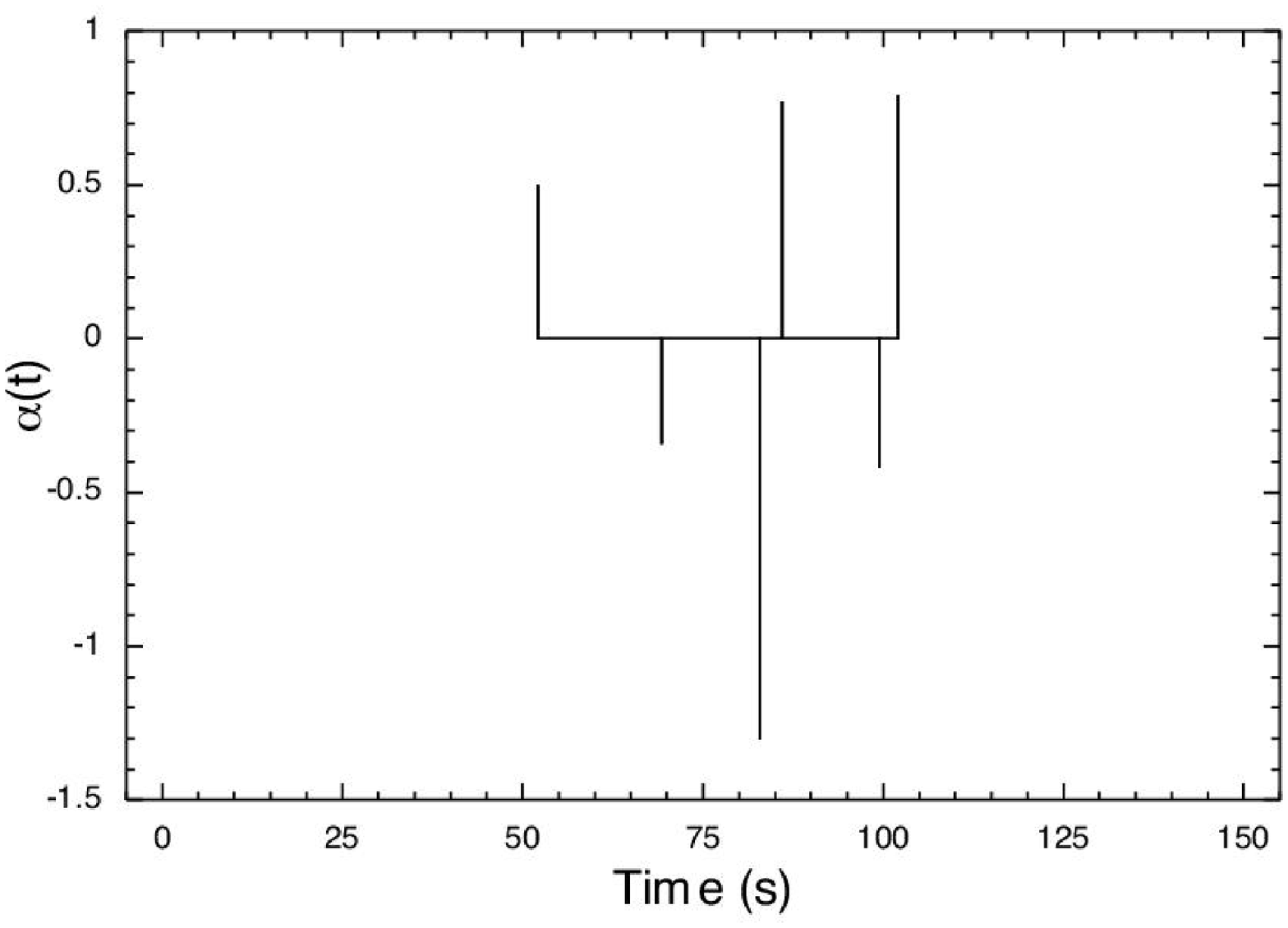}
\includegraphics[width=3.7 in, angle=0.0]{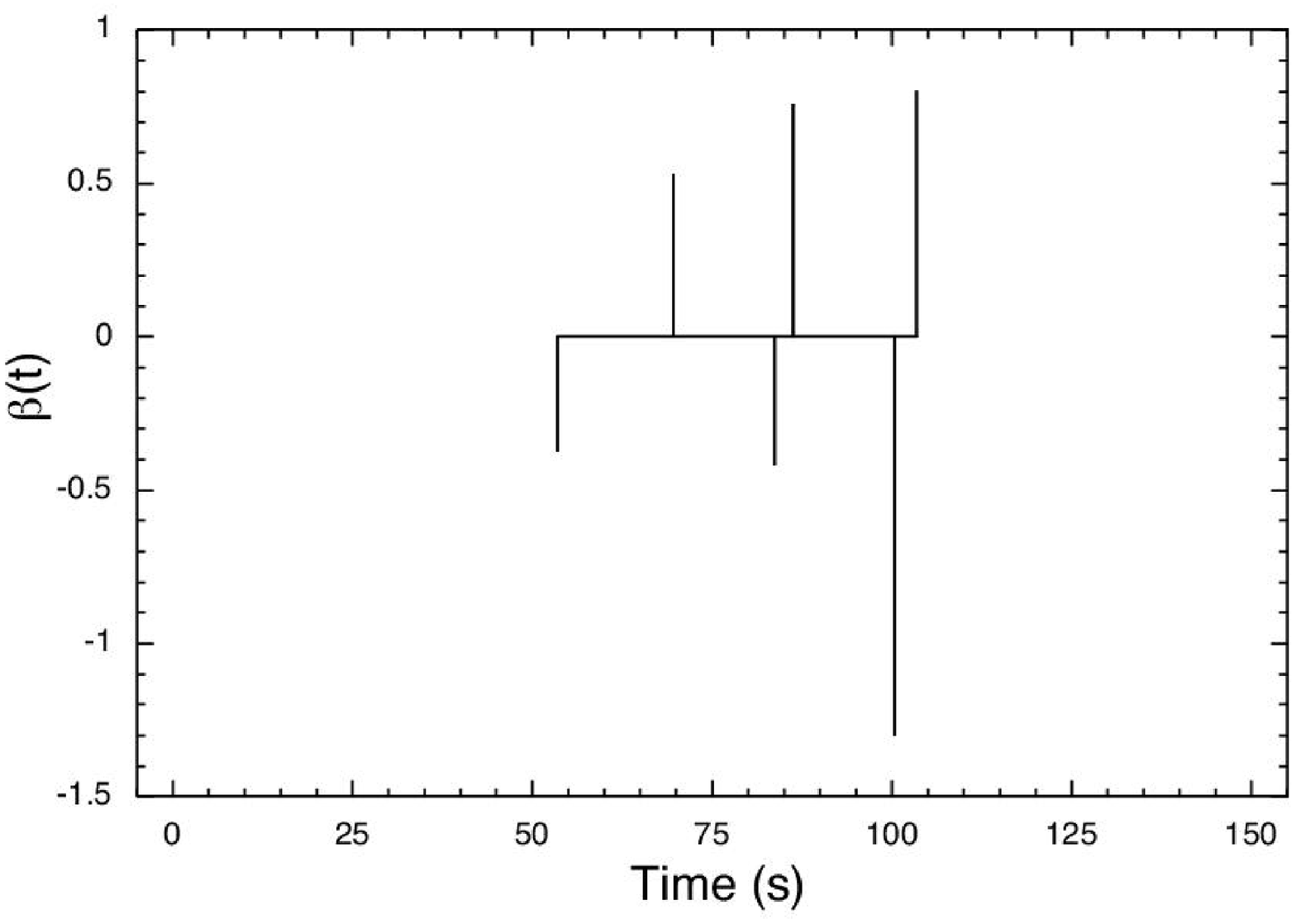}
\includegraphics[width=3.7 in, angle=0.0]{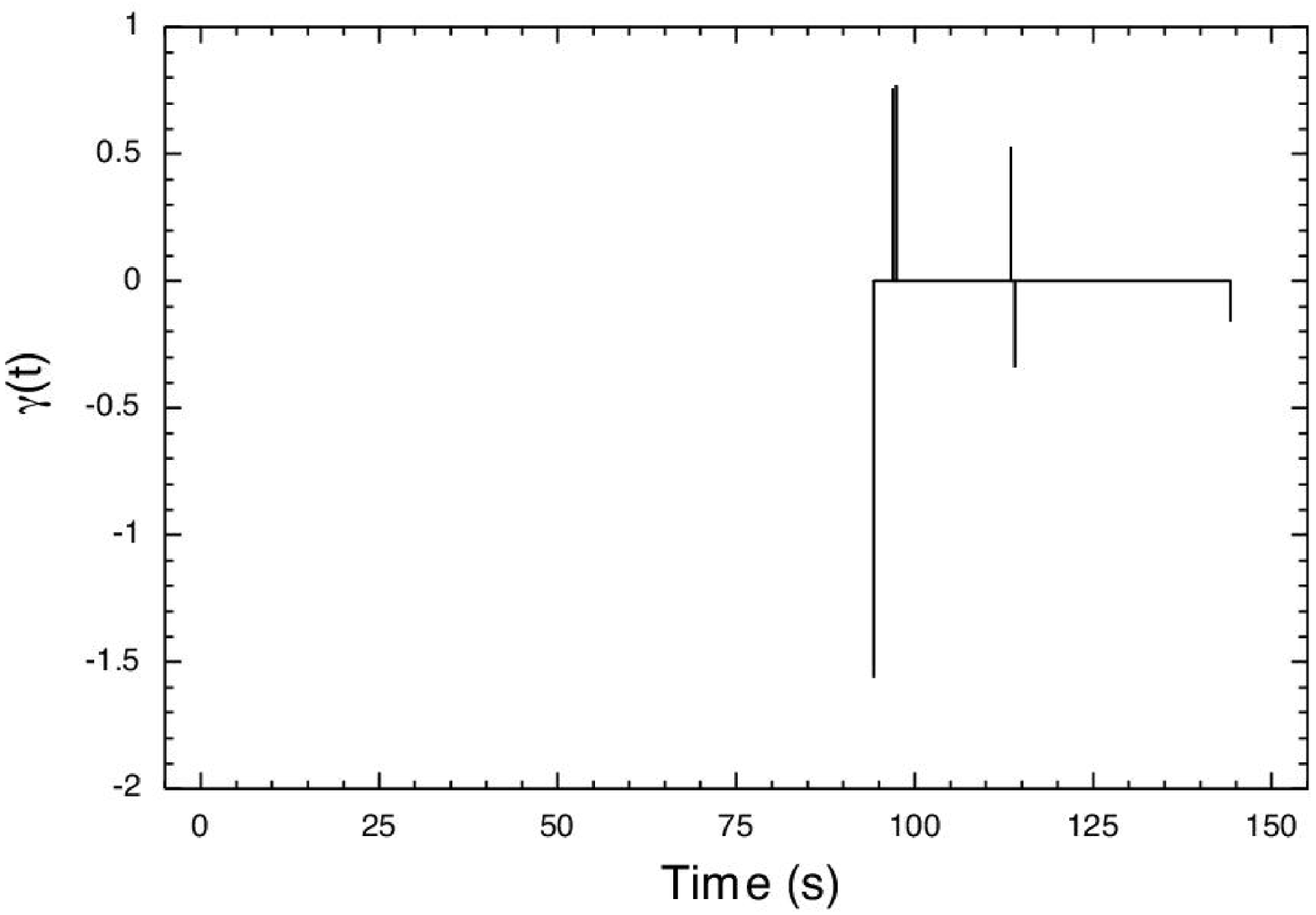}
\caption{The responses of the three Sagnac combinations,
  $\alpha$ (a), $\beta$ (b), and $\gamma (c)$, to a gravitational wave
  burst of unit amplitudes. The source has been assumed to be located
  at: $\theta_s = 32^\circ \ , \ \phi_s = 107^\circ$.}
\end{figure}

\begin{figure}
 \centering
\includegraphics[width=3.3 in, angle=0.0]{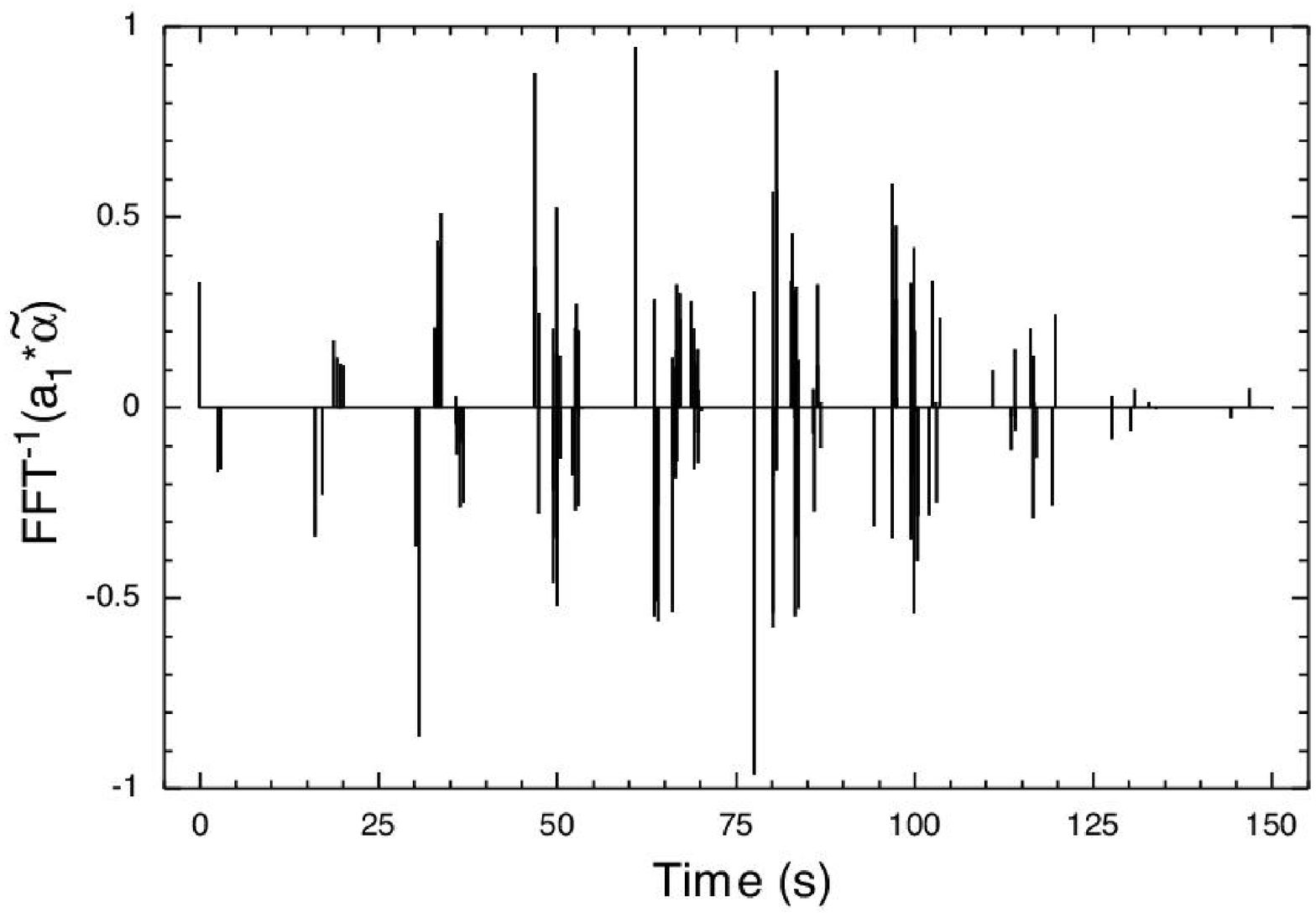}
\includegraphics[width=3.3 in, angle=0.0]{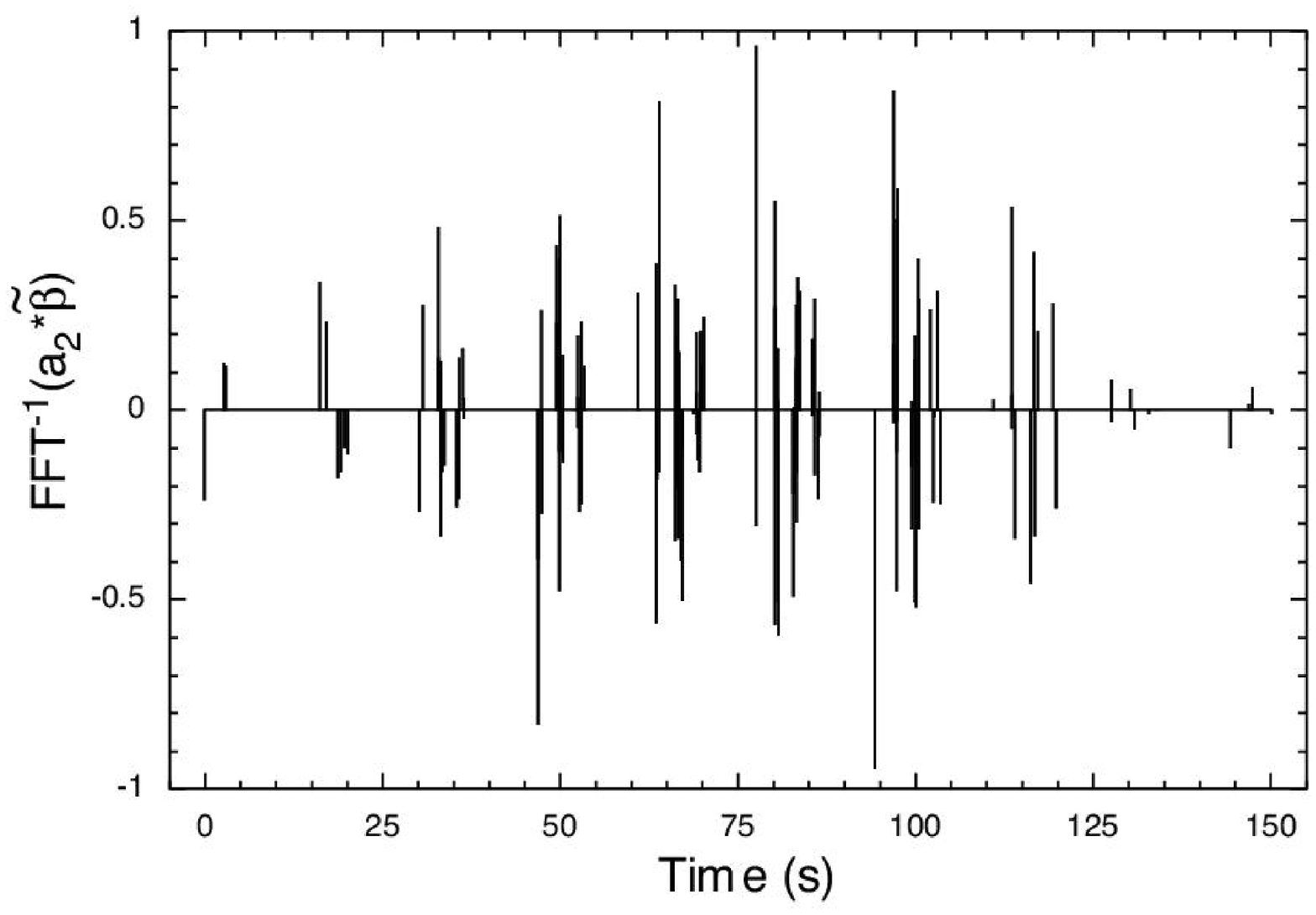}
\includegraphics[width=3.3 in, angle=0.0]{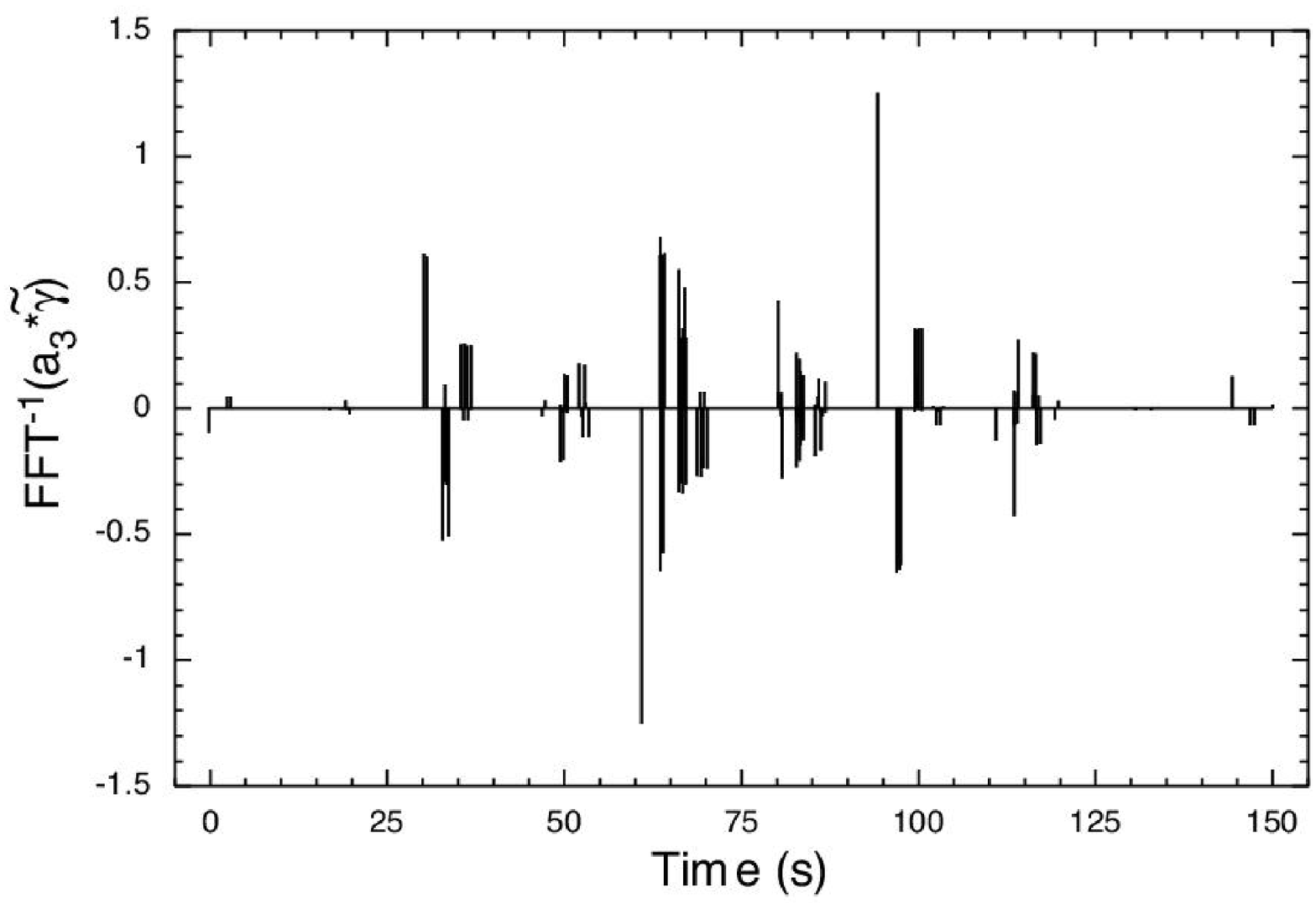}
\includegraphics[width=3.3 in, angle=0.0]{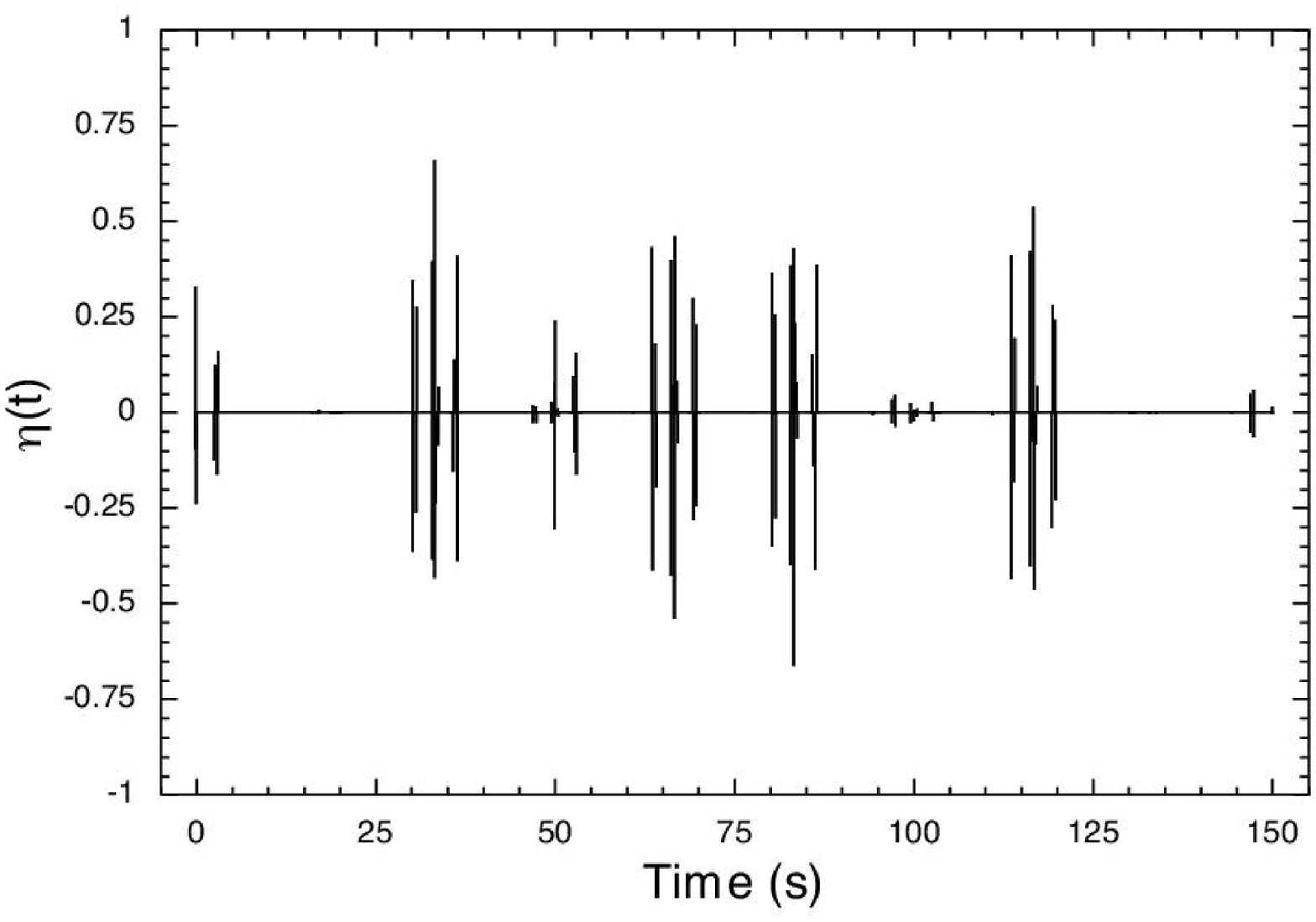}
\caption{The time domain responses
  of the Inverse-Fourier Transforms of the three functions, $a_1 \ 
  {\widetilde \alpha}$ (a), $a_2 \ {\widetilde \beta}$ (b), $a_3 \ 
  {\widetilde \gamma}$ (c), which determine the ZSS. The angles
  ($\theta, \phi$), over which the search is performed have been fixed
  to be equal to $\theta = 31.5^\circ \ , \ \phi = 106.5^\circ$; (d)
  the ZSS constructed from the functions shown in (a, b, c).}
\end{figure}

\begin{figure}
 \centering
\includegraphics[width=3.4 in, angle=0.0]{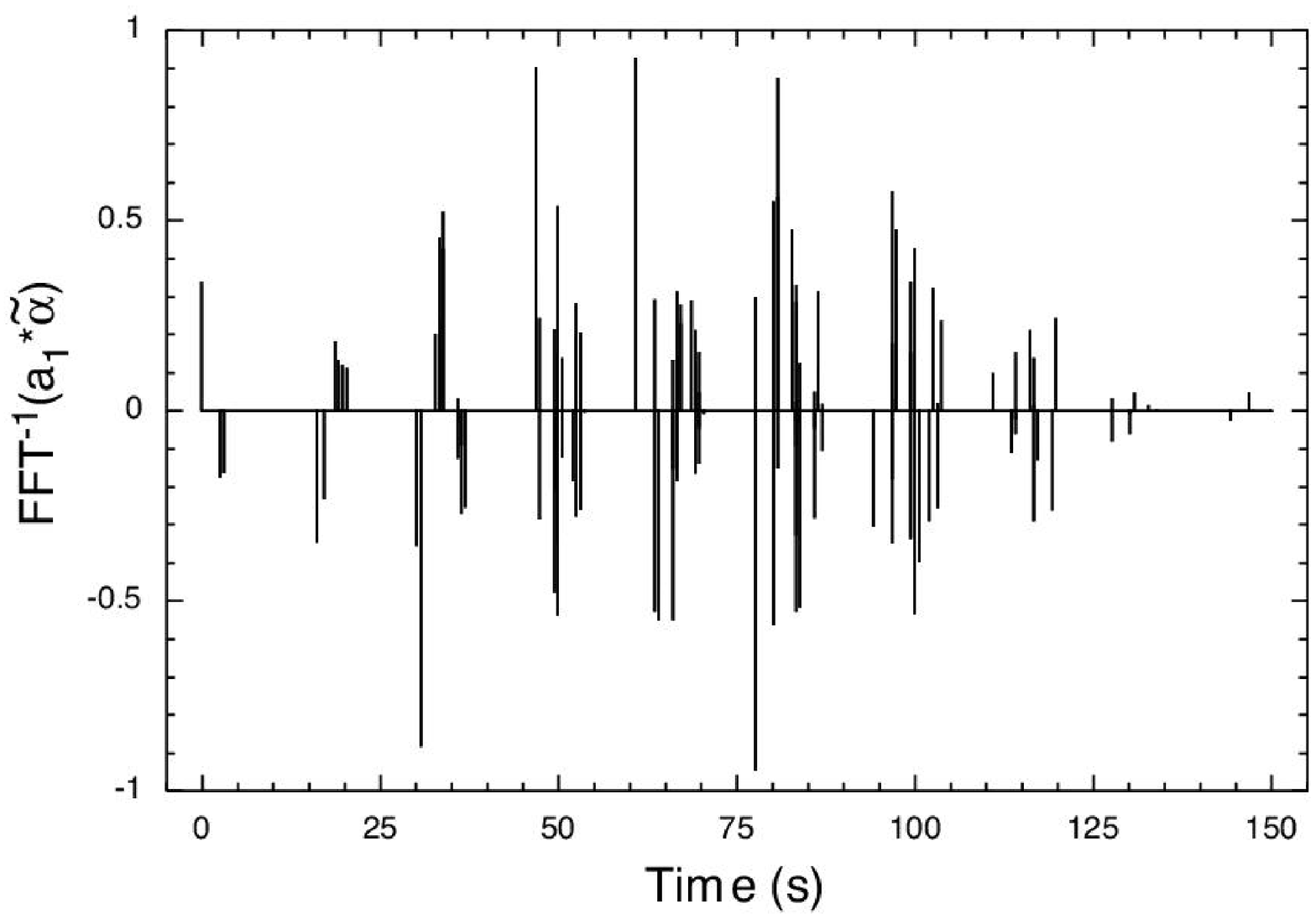}
\includegraphics[width=3.4 in, angle=0.0]{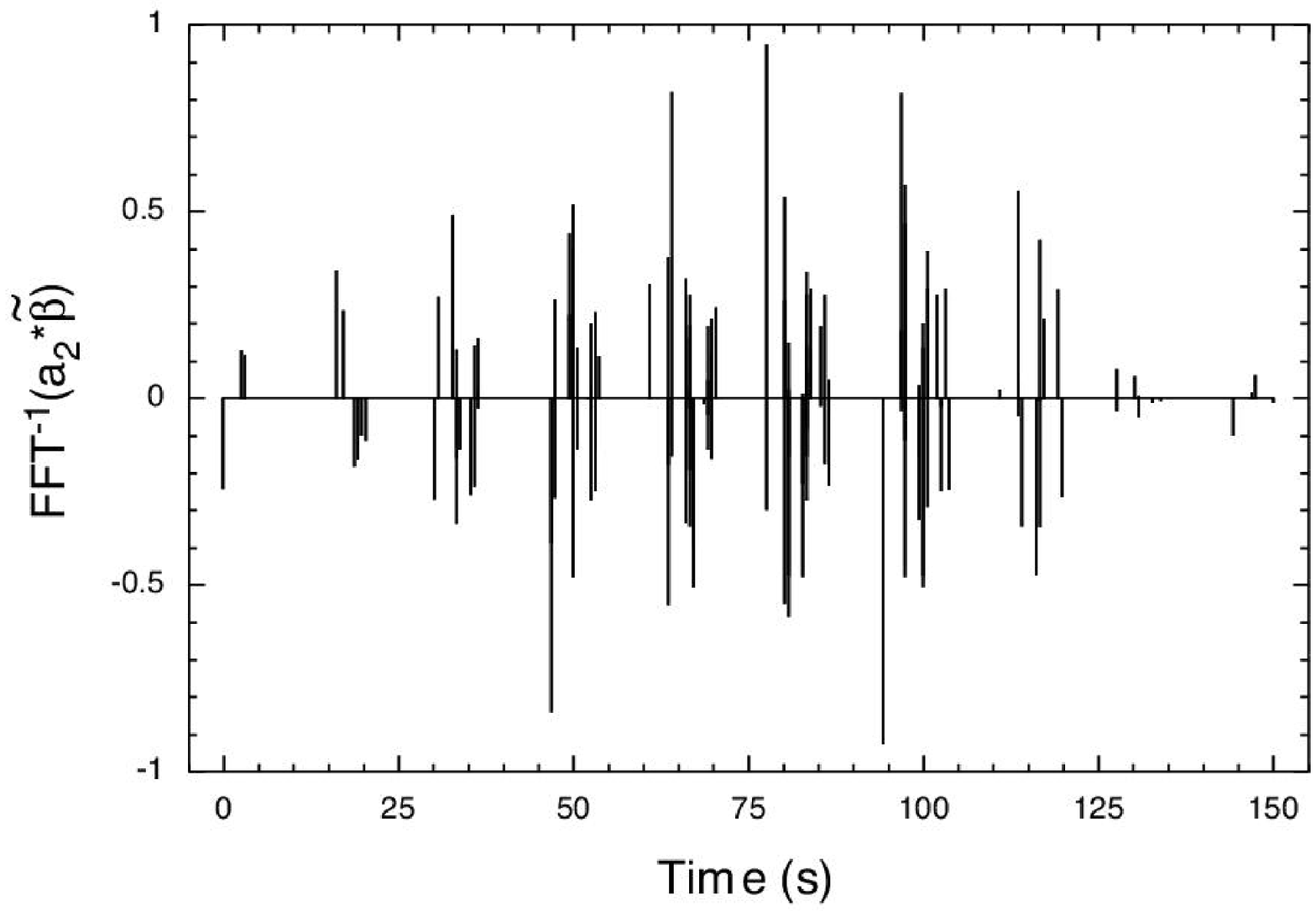}
\includegraphics[width=3.4 in, angle=0.0]{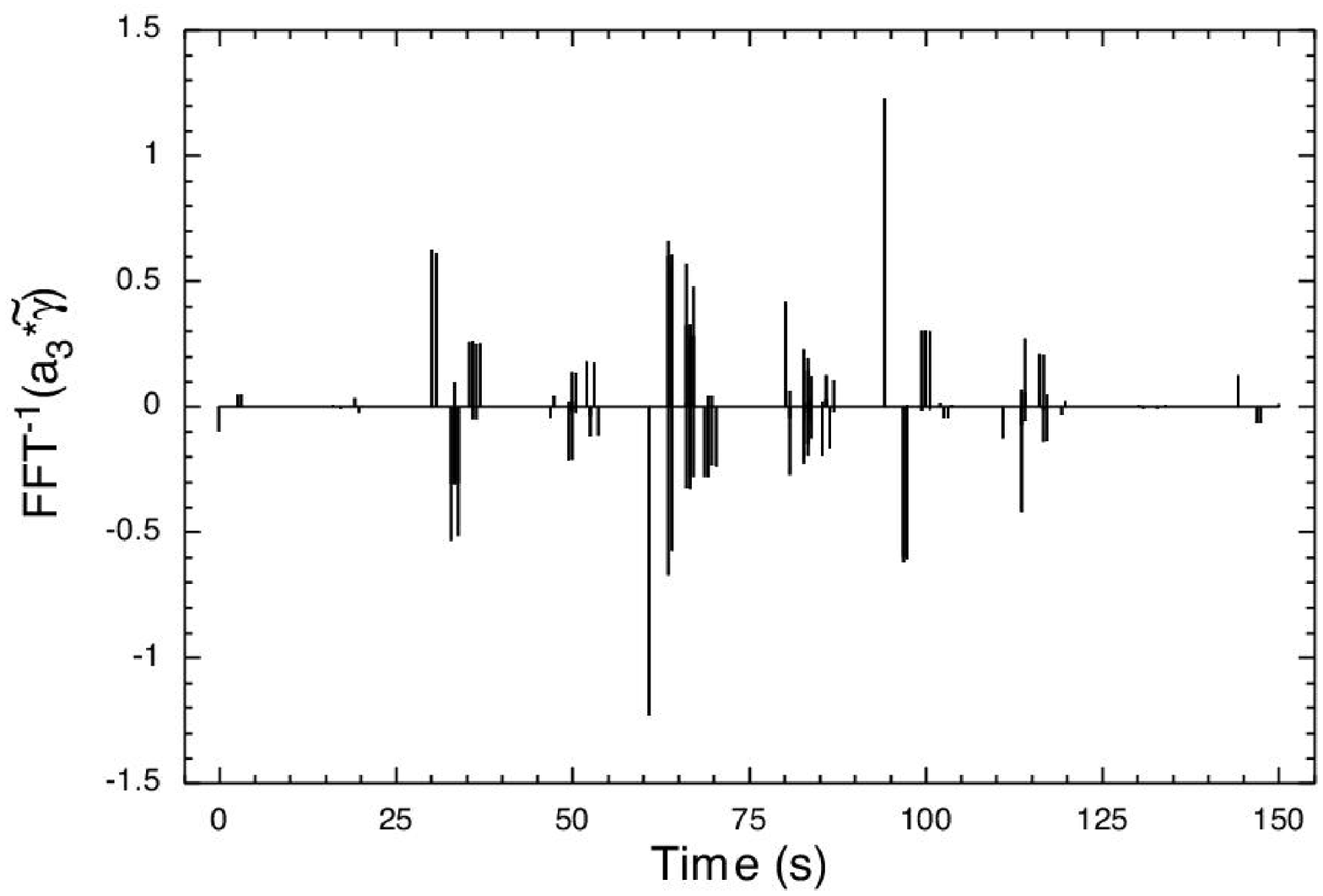}
\includegraphics[width=3.4 in, angle=0.0]{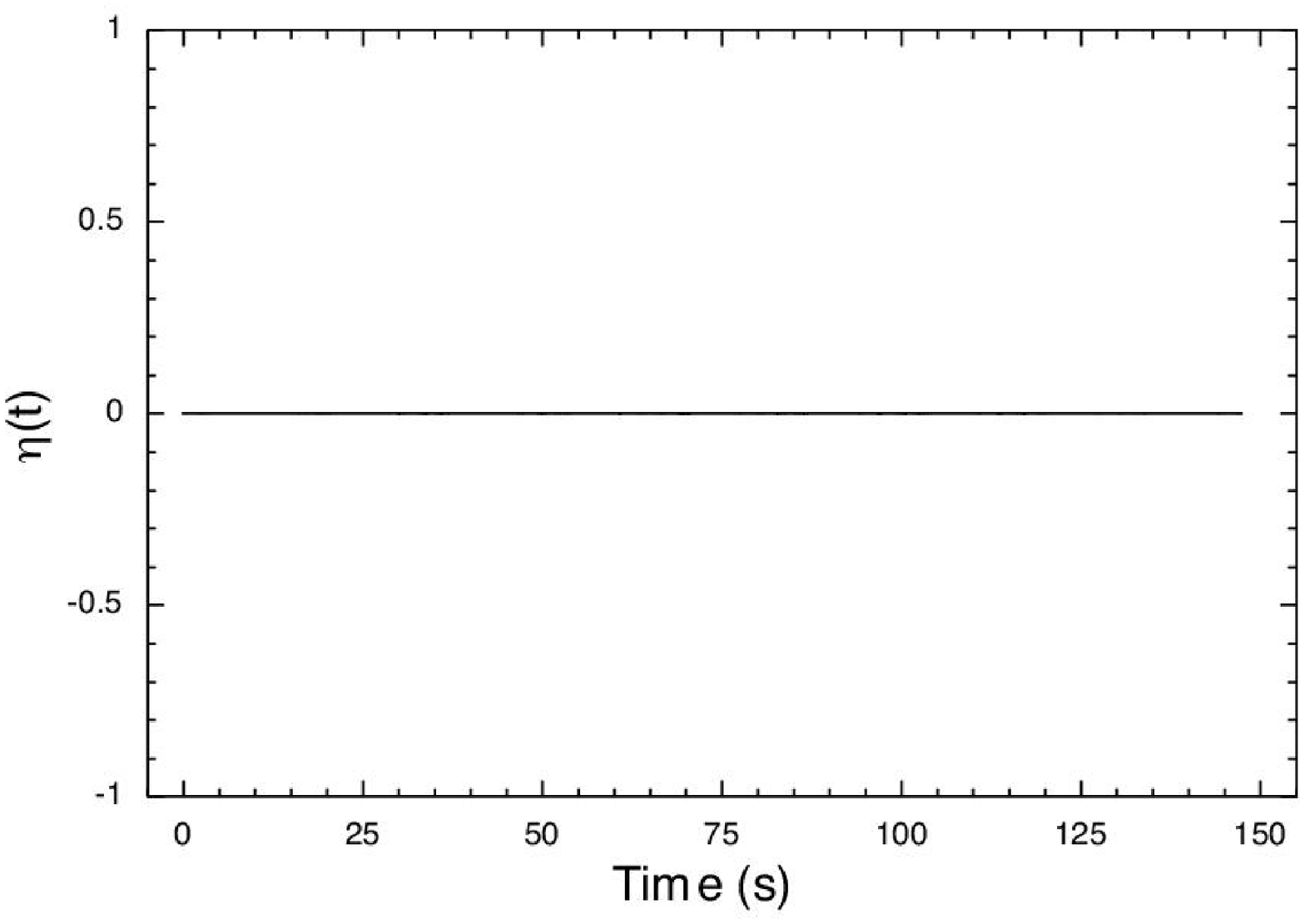}
\caption{Plots (a, b, c, d) are as in Figure 6, but now with ($\theta
  = \theta_s \ , \ \phi = \phi_s$).}
\end{figure}
Figures (5, 6, 7) schematically show the implementation of the ZSS in the case
of a gravitational wave burst whose wave components, $h_+ (t)$,
$h_\times (t)$, are delta-like pulses of unit amplitudes.  The
noiseless responses of the three Sagnac combinations to such a
gravitational wave signal, which we have assumed to be incoming from
($\theta_s = 32^\circ, \phi_s = 107^\circ$), are given in Figure 5 (a, b, c).
Note the characteristic six-pulse structure of their responses. In
Figures 6 (a, b, c) the time-domain expressions of the inverse Fourier
transforms of the functions $a_1 (f, \theta, \phi) \ 
{\widetilde{\alpha}} (f)$, $a_2 (f, \theta, \phi) \ 
{\widetilde{\beta}} (f)$, $a_3 (f, \theta, \phi) \ 
{\widetilde{\gamma}} (f)$ are plotted for values of the angles
($\theta = 31.5^\circ , \phi = 106.5^\circ$). The corresponding ZSS
combination is given in Figure 6 (d), showing that no cancellation is
achieved for values of the polar angles slightly different from those
corresponding to the source location. Figure 7 (a, b, c, d) shows
instead plots similar to those given by Figure 6, but now
with $(\theta , \phi) = (\theta_s, \phi_s)$.

\section{The Long-Wavelength Limit}

In order to understand the behavior of the ZSS, $\eta$, in the
low-part of the LISA frequency band, we provide in this section its
analytic expression in the long-wavelength limit (LWL), i.e.  when the
characteristic wavelength of the GW signal, $\lambda$, is much larger
than the typical scale size of the array, $L$.

Let us rewrite the expression for $\eta$ in the Fourier domain (see
equation (\ref{eq:13}) above)
\begin{eqnarray}
{\widetilde \eta} & \equiv & [\beta_+ (f, \theta, \phi) \
\gamma_\times (f, \theta, \phi) \ - \ \beta_\times (f, \theta,
\phi) \ \gamma_+ (f, \theta, \phi)] \ {\widetilde{\alpha}} (f) 
\nonumber
\\
& + & [\gamma_+ (f, \theta, \phi) \
\alpha_\times (f, \theta, \phi) \ - \ \gamma_\times (f, \theta,
\phi) \ \alpha_+ (f, \theta, \phi)] \ {\widetilde{\beta}} (f) 
\nonumber
\\
& + & [\alpha_+ (f, \theta, \phi) \
\beta_\times (f, \theta, \phi) \ - \ \alpha_\times (f, \theta,
\phi) \ \beta_+ (f, \theta, \phi)] \ {\widetilde{\gamma}} (f) \ ,
\label{eq:13c}
\end{eqnarray}
where the functions $\alpha_+$, $\alpha_\times$, $\beta_+$,
$\beta_\times$, and $\gamma_+$, $\gamma_\times$ are given in appendix
B. We have been able to find the long-wavelength expansion ($fL <<1$)
of ${\widetilde \eta}$ by using the program {\it Mathematica}
\cite{Wolf02}. Since the LISA arm lengths will differ at most by a few
percents \cite{PPA98}, the general expression of the ZSS in the long
wavelength limit can be written in the following simpler form
\begin{eqnarray}
{\widetilde \eta} (f, \theta, \phi) & \approx & (2 \ \pi f \ L)^4 \ 
[(F_1^+ F_2^\times - F_1^\times F_2^+) +
(F_2^+ F_3^\times - F_2^\times F_3^+) +
(F_3^+ F_1^\times - F_3^\times F_1^+)]
\nonumber
\\
& & \times \left\{ \frac{i \ (2 \ \pi \ f \ L)^3}{4} \ ({\hat k}^{(s)} -
{\hat k}) 
\cdot \left[{\hat n_1} \cdot {\widetilde {\bf h^{(s)}}}(f) \cdot {\hat n_1} +
{\hat n_2} \cdot {\widetilde {\bf h^{(s)}}}(f) \cdot {\hat n_2}
\right. \right.
\nonumber
\\
& & \left. \left. + {\hat n_3} \cdot {\widetilde {\bf h^{(s)}}}(f) \cdot {\hat n_3}\right]
+ {\widetilde \alpha_n} + {\widetilde \beta_n} + {\widetilde \gamma_n} \right\} \ ,
\label{eq:13d}
\end{eqnarray}
where $L$ is the nominal LISA arm length ($\simeq 5 \ \times 10^6$
km), the label $^{(s)}$ on a specific functions means ``evaluated at
$\theta = \theta_s \ ; \ \phi = \phi_s$ '', and (${\widetilde
  \alpha_n}, {\widetilde \beta_n}, {\widetilde \gamma_n}$) are the
Fourier transforms of the three random processes associated with the
noises in ($\alpha, \beta, \gamma$).  Note that as $(k_x, k_y, \pm
k_z) \to (k_x, k_y, \pm k_z)^{(s)}$, ${\widetilde \eta} \to 0$ as expected. This is
because the unit vector $\hat k$, corresponding to a point that is the
mirror image of the source location relative to the LISA plane, will
have the same scalar product with the three unit vectors ${\hat n}_j$
as ${\hat k}^{(s)}$.

It is interesting to compare the LWL expression of the ZSS given
above, against that for the Symmetric Sagnac combination, $\zeta$
\cite{TAE} 
\begin{eqnarray}
{\widetilde{\zeta}}^{gw} (f) & \simeq & {1 \over {12}} \ (2 \pi i f L)^3 \ 
\left[
({\hat k} \cdot {\hat n_1}) ({{\hat n_1} \cdot {\widetilde{{\bf h}}}(f) \cdot {\hat n_1}}) +
({\hat k} \cdot {\hat n_2}) ({{\hat n_2} \cdot {\widetilde{{\bf h}}}(f) \cdot {\hat n_2}}) \right. 
\nonumber \\
& & \ \ \ \ \ \ \ \ \ \ \ \ \ \ \ \ \ \ 
+
\left.
({\hat k} \cdot {\hat n_3}) ({{\hat n_3} \cdot {\widetilde{{\bf h}}}(f) \cdot {\hat n_3}}) 
\right] + \frac{1}{3} 
\left[{\widetilde \alpha_n} + {\widetilde \beta_n} + {\widetilde
    \gamma_n} \right] \ ,
\label{eq:13e}
\end{eqnarray}
where the noise affecting $\zeta$ can be rewritten (in the LWL) in
terms of the ($\alpha, \beta, \gamma$) noises by using the LWL limit
of the Fourier transform of equation (\ref{eq:6}) with equal arm
lengths. Although both combinations display the $(2 \pi f
L)^3$-dependence in the signal and the same noises, the ability of
${\widetilde \eta}$ to further suppress the gravitational wave signal by
identifying the direction it came from should be regarded as an
improvement over $\zeta$. This property will be particularly useful
for discriminating exceptionally strong signals against random noise
fluctuations that will be still observable by $\zeta$ in this
low-part of the LISA frequency band.

\section{Summary and Conclusions}

We have identified a method for nulling a gravitational wave signal
observed in coincidence by three LISA TDI combinations.  We have named
the resulting TDI combination the {\it Zero-Signal Solution}.  The
method can in principle identify two points in the sky where a
gravitational wave burst, or a sinusoidal signal, came from. In order
for the method to work in the case of a sinusoidal signal, the
gravitational wave signal must be observable after integrating the
data over a period such that the Doppler modulation induced by the
LISA motion around the Sun can be disregarded.

Although the analytic expression of the ZSS has been derived in terms
of the three Sagnac observables ($\alpha, \beta, \gamma$), any other
three TDI combinations capable of generating the entire TDI space
(such as the Unequal-Arm Michelson combinations $X, Y, Z$) could have
been used. In particular, the TDI combinations ($A, E, T$)
\cite{Prince02}, whose noises are uncorrelated, would have worked as
well. The use of the ($A, E, T$) combinations will be particularly
useful when we will estimate the accuracies provided by the ZSS in
determining the source location and the wave's two amplitudes. These
points will be addressed in our follow up article.

\begin{acknowledgments}
  SLL acknowledges support from LISA contract number PO 1217163. The
  research was performed at the Jet Propulsion Laboratory, California
  Institute of Technology, under contract with the National
  Aeronautics and Space Administration.
\end{acknowledgments}

\appendix
\section{The Beam Pattern Functions}

This appendix derives the analytic expressions of the antenna patterns
$F_j^+ (\theta, \phi)$, $F_j^\times (\theta, \phi)$ entering in the
wave functions $\Psi_j (t) \ , \ j=1, 2, 3$, and in the 
pattern functions $\alpha_{+}$, $\alpha_{\times}$, $\beta_{+}$,
$\beta_{\times}$, $\gamma_{+}$, $\gamma_{\times}$ introduced in
equations (\ref{eq:9a}, \ref{eq:9b}, \ref{eq:9c}), and given in
appendix B.

Consider a plane gravitational wave of direction of propagation $\hat
k$ incident on the LISA spacecraft constellation.  Let the wave
coordinate system be ($x', y', z'$), with the wave traveling in the
$z'$ direction, and let the axes with respect to which $h_+ (t)$, and
$h_{\times} (t)$ are referred to be the ($x', y'$) axes. The
coordinate system associated with LISA, ($x, y, z$) is obtained from
them by a rotation described by the polar angles ($\theta, \phi$)
(Figure 3). For sake of simplicity, and without lack of generality, we
can assume the $x'$ axis to be parallel to the ($x, y$) plane, and the
wave's two independent amplitudes $h_+ (t)$, and $h_{\times} (t)$,
associated with the two independent polarizations, to be defined with
respect to this choice of coordinates.

Let us now consider the null vector $\vec m$ defined by
\begin{equation}
{\vec m} \equiv \frac{1}{\sqrt{2}} ({\hat e_{x'}} + i \ {\hat e_{y'}}) \ ,
\end{equation}
where $\hat e_{x'}$, $\hat e_{y'}$ are the unit vectors in the $x'$, $y'$
directions, respectively. The tensor $\bf h$ is then just \cite{GT}
\begin{equation}
{\bf h} = 2 \ h_+ (t) \ Re({\vec m} \otimes {\vec m}) + 
2 \ h_\times (t) \ Im ({\vec m} \otimes {\vec m}) \ .
\label{eq:A1}
\end{equation}
In the LISA coordinate system the components of $h_{ij}$ can be
obtained by finding the components of $\vec m$ in these coordinates.
The vector $\vec m$ can be written as
\begin{equation}
{\vec m} = \frac{1}{\sqrt{2}} \  [(- \sin\phi + i \ \cos\theta \cos\phi) \
{\hat e_x} + (\cos\phi + i \ \cos\theta \ \sin\phi)
\ {\hat e_y} + (i \ \sin\theta) \ {\hat e_z}] \ , 
\label{eq:A2}
\end{equation}
while the unit vector $\hat k$, describing the direction of
propagation of the wave from the source to LISA assumes the following
form
\begin{equation}
{\hat k} = (- \sin\theta \ \cos\phi) \ {\hat e_x} + (- \sin\theta \
\sin\phi ) \ {\hat e_y} + (- \cos\theta) \ {\hat e_z} \ .
\label{eq:A3}
\end{equation}
Since the $\Psi_j (t) \ , \ j=1, 2, 3$ functions, entering into the
expressions of $\alpha^{gw}$,  $\beta^{gw}$,  $\gamma^{gw}$ have the
following analytic forms
\begin{eqnarray} 
\Psi_j (t) & = & {1 \over 2} \ 
  \ {{{\hat n_j} \cdot {\bf h}(t) \cdot {\hat n_j}} \over { 1 - ({\hat
        k} \cdot {\hat n_j})^2}} \ , 
\label{eq:A4}
\end{eqnarray} 
after replacing equation (\ref{eq:A1}) into (\ref{eq:A4}), and
performing some simple algebra, we can rewrite $\Psi_j (t)$ in the
following form
\begin{equation}
\Psi_j (t) \equiv \frac{F_j^+ (\theta, \phi) \ h_+ (t) + 
F_j^\times (\theta, \phi) \ h_\times (t)}
{2 \ ({ 1 - ({\hat k} \cdot {\hat n_j})^2})}
\end{equation}
where the functions $F_j^+ (\theta, \phi)$,
$F_j^\times (\theta, \phi)$ are equal to
\begin{eqnarray}
F_j^+ (\theta, \phi) & \equiv & 
2 \ [{{\hat n_j} \cdot Re({\vec m} \otimes {\vec m}) \cdot {\hat n_j}}]
\ , 
\label{eq:A5a}
\\
F_j^\times (\theta, \phi) & \equiv & 
2 \ [{{\hat n_j} \cdot Im({\vec m} \otimes {\vec m}) \cdot {\hat n_j}}]
\ , 
\label{eq:A5b}
\end{eqnarray}
If we now rewrite the vector $\vec m$ in the following form
\begin{equation}
{\vec m} \equiv {\vec m}_1 + i \ {\vec m}_2 \ ,
\label{eq:A6}
\end{equation}
where the two vectors ${\vec m}_1$, ${\vec m}_2$ have components
\begin{eqnarray}
{\vec m}_1 & = & \frac{1}{\sqrt{2}} \ 
[(-\sin\phi) \ {\hat e_x} + 
(\cos\phi ) \ {\hat e_y} ] \ ,
\nonumber
\label{eq:A7a}
\\
{\vec m}_2 & = & \frac{1}{\sqrt{2}} \ 
[(\cos\theta \ \cos\phi) \ {\hat e_x} + 
(\cos\theta \ \sin\phi) \ {\hat e_y}
+ (\sin\theta) \ {\hat e_z}] \ .
\label{eq:A7b}
\end{eqnarray}
the tensor product ${\vec m} \otimes {\vec m}$ becomes equal to
\begin{equation}
{\vec m} \otimes {\vec m} = 
[({\vec m}_1 \otimes {\vec m}_1) - ({\vec m}_2 \otimes {\vec m}_2)] + i \ 
[({\vec m}_1 \otimes {\vec m}_2) + ({\vec m}_2 \otimes {\vec m}_1)] \ .
\end{equation}
The expressions for $F_j^+$, $F_j^\times$ given in equations
(\ref{eq:A5a}, \ref{eq:A5b}) can be rewritten as
\begin{eqnarray}
F_j^+ & = & 2 \ [({\hat n_j} \cdot {\vec m}_1)^2 - 
({\hat n_j} \cdot {\vec m}_2)^2 ] \ ,
\label{eq:A8a}
\\
F_j^\times & = & 
4 \ ({\hat n_j} \cdot {\vec m}_1) \ ({\hat n_j} \cdot {\vec m}_2) 
\ , \ j=1, 2, 3 \ .
\label{eq:A8b}
\end{eqnarray}
Since the unit vector $\hat n_j$ are in the ($x, y$)-plane, they can
be expressed as
\begin{equation}
\hat n_j = \cos(\nu_j) \ {\hat e_x} + \sin(\nu_j) \ {\hat e_y} 
\ , \ j=1, 2, 3 \ ,
\label{eq:A9}
\end{equation}
and equations (\ref{eq:A8a}, \ref{eq:A8b}), after some simple algebra,
assume the following form
\begin{eqnarray}
F_j^+ & = & [\sin^2(\nu_j - \phi) - \cos^2(\nu_j - \phi) \ \cos^2\theta] \ ,
\label{eq:A10a}
\\
F_j^\times & = & [\cos\theta \ \sin2(\nu_j - \phi)] \ , \ j=1, 2, 3 \ .
\label{eq:A10b}
\end{eqnarray}
Equations (\ref{eq:A10a}, \ref{eq:A10b}) show that the $F_j^+$
beam-pattern functions are invariant under the transformation $\theta
\to \pi - \theta$, while the $F_j^\times$ change sign. These symmetry
properties of the $F_j^{+,\times}$ functions imply that the ZSS changes
sign under the same transformation, an important property highlighted
in Section II.

\section{The ZSS coefficients $a_1$, $a_2$, $a_3$}

In order to derive the expression of the ZSS in the Fourier domain we
have introduced the {\it generalized beam-pattern functions}
$\alpha_+, \alpha_\times$, $\beta_+, \beta_\times$, and $\gamma_+,
\gamma_\times$ (see equations \ref{eq:9a}, \ref{eq:9b}, \ref{eq:9c}),
while the corresponding expression in the time-domain (equation
\ref{eq:14}) was written in terms of eighty-one amplitudes,
$A_j^{(k)}$, and corresponding time-delays, $\Delta_j^{(k)} \ , \ j=1,
2, 3 \ , \ k=1, 2, ..., 27$. In this appendix we provide the analytic
expressions of these functions. 

Using the identities $\mu_{i} = \vec{k}\cdot\hat{p_{i}}$ and
\begin{equation}
    w_{1} \equiv \hat{k}\cdot\hat{n}_{1} = \frac{\ell}{L_{1}} (\mu_{2} -
    \mu_{3}) \ ,
    \label{w1}
\end{equation}
\begin{equation}
    w_{2} \equiv \hat{k}\cdot\hat{n}_{2} = \frac{\ell}{L_{2}} (\mu_{3} -
    \mu_{1}) \ ,
    \label{w2}
\end{equation}
\begin{equation}
    w_{3} \equiv \hat{k}\cdot\hat{n}_{3} = \frac{\ell}{L_{3}} (\mu_{1} -
    \mu_{2}) \ ,
    \label{w3}
\end{equation}
the signal part of $\alpha(t)$ is written in the time domain as
\begin{eqnarray}
     \alpha^{gw} & = &
     \left[1 - w_{2}\right] \left[\Psi_{2}(t - L_{2} - \mu_{3}\ell)
          - \Psi_{2}(t - \mu_{1}\ell)\right] \nonumber \\
     & - & \left[1 + w_{2}\right] \left[\Psi_{2}(t - L_{1} - L_{2} - 
            L_{3} - \mu_{1}\ell) - \Psi_{2}(t - L_{1}
            - L_{3} - \mu_{3}\ell) \right] \nonumber \\
     & + & \left[1 - w_{3}\right] \left[\Psi_{3}(t - L_{1} - L_{2} -
            L_{3} - \mu_{1}\ell) - \Psi_{3}(t - L_{1} - L_{2} - 
            \mu_{2}\ell)\right] \nonumber \\
     & - & \left[1 + w_{3}\right] \left[\Psi_{3}(t - L_{3} - 
           \mu_{2}\ell) - \Psi_{3}(t - \mu_{1}\ell)\right] \nonumber \\
     & + & \left[1 - w_{1}\right] \left[\Psi_{1}(t - L_{1} - L_{2} - 
           \mu_{2}\ell) - \Psi_{1}(t - L_{2} - \mu_{3}\ell)\right] 
           \nonumber \\
     & - & \left[1 + w_{1}\right] \left[\Psi_{1}(t - L_{1} - L_{3} - 
     \mu_{3}\ell) - \Psi_{1}(t - L_{3} - \mu_{2}\ell)\right]\ .
     \label{AlphaGW}
\end{eqnarray}

Since in the frequency domain the functions $\Psi_i$ are equal to
\begin{equation}
    \tilde{\Psi}_{i} = \frac{F_{i}^{+} \tilde{h}_{+} + F_{i}^{\times}
    \tilde{h}_{\times}}{2(1 - w_{i}^{2})}\ ,
    \label{PsiDefn}
\end{equation}
the response of the interferometric variable $\alpha$, to a
gravitational wave signal, becomes
\begin{equation}
    \tilde{\alpha}(f) = \alpha_{+}(f, \theta, \phi) \tilde{h}_{+} + 
    \alpha_{\times}(f, \theta, \phi) \tilde{h}_{\times}\ .
    \label{alphaFactored}
\end{equation}
where the coefficients $\alpha_{+,\times}(f, \theta, \phi)$ may be
written as
\begin{equation}
   \alpha_{+,\times}(f, \theta, \phi) =
        \alpha_{1} F_{1}^{+,\times} +
        \alpha_{2} F_{2}^{+,\times} +
        \alpha_{3} F_{3}^{+,\times} \ .
   \label{alphaPlusCross}
\end{equation}
In equation (\ref{alphaPlusCross}) the functions $\alpha_1$,
$\alpha_2$, $\alpha_3$ are given by the following expressions
\begin{equation}
   \alpha_{1} = \frac{(e^{i 2 \pi f(L_{1} + L_{2} + \mu_{2}\ell)}
    - e^{i 2 \pi f(L_{2} + \mu_{3}\ell)})}{2(1 + w_{1})}
     + \frac{(e^{i 2 \pi f(L_{3} + \mu_{2}\ell)} -
    e^{i 2 \pi f(L_{1} + L_{3} + \mu_{3}\ell)})}{2(1 - w_{1})}
    \label{alpha1}
\end{equation}

\begin{equation}
   \alpha_{2} = \frac{(e^{i 2 \pi f(L_{2} + \mu_{3}\ell)} - e^{i 2 \pi
   f(\mu_{1}\ell)})}{2(1 + w_{2})} +
   \frac{(e^{i 2 \pi f(L_{1} + L_{3} + \mu_{3}\ell)} -
    e^{i 2 \pi f(L_{1} + L_{2} + L_{3} + \mu_{1}\ell)})}{2(1 - w_{2})}
    \label{alpha2}
\end{equation}

\begin{equation}
    \alpha_{3} = \frac{(e^{i 2 \pi f(L_{1} + L_{2} + L_{3} +
    \mu_{1}\ell)} - e^{i 2 \pi f(L_{1} + L_{2} + \mu_{2}\ell)})}{2(1 +
    w_{3})} + \frac{(e^{i 2 \pi f(\mu_{1}\ell)} - e^{i 2 \pi f(L_{3} +
    \mu_{2}\ell)})}{2(1 - w_{3})}
    \label{alpha33}
\end{equation}
The numerical indices are again spacecraft indices, which can be
permuted in order to obtain the corresponding functions for the other
TDI combinations $\beta$, and $\gamma$ (indices $1 \rightarrow 2
\rightarrow 3 \rightarrow 1$, such that $\alpha \rightarrow \beta
\rightarrow \gamma \rightarrow \alpha$).

The $a_{3}$ coefficient in the Zero Signal Solution is built from 
combinations of the pattern functions that determine
the responses of $\tilde{\alpha}$ and $\tilde{\beta}$ to the
gravitational wave signal. Its analytic expression can now be
rewritten as
\begin{eqnarray}
   a_{3} & = & \alpha_{+}\beta_{\times} - \alpha_{\times}\beta_{+} 
   \nonumber \\
   & = &
   (\alpha_{1} \beta_{2} - \alpha_{2} \beta_{1})
   (F_{1}^{+}F_{2}^{\times} - F_{1}^{\times}F_{2}^{+}) \nonumber \\
   & + & (\alpha_{3} \beta_{1} - \alpha_{1} \beta_{3})
   (F_{3}^{+}F_{1}^{\times} - F_{3}^{\times}F_{1}^{+}) \nonumber \\
   & + & (\alpha_{2} \beta_{3} - \alpha_{3} \beta_{2})
   (F_{2}^{+}F_{3}^{\times} - F_{2}^{\times}F_{3}^{+})
   \ .
   \label{combo1}
\end{eqnarray}

After algebraically expanding the expression above
we can rewrite $a_{3}$ in the following form
\begin{equation}
a_{3} = \sum_{k=1}^{27} A_{3}^{(k)} \ e^{2 \pi if \Delta_{3}^{(k)}}
\ ,
\label{expand1}
\end{equation}
where the $\Delta_{3}^{(k)}$ are $27$ different time shifts associated
with the coefficient $a_3$, and the $A_3^{a}$ are functions that only
depend on the geometric angles $(\theta,\phi)$. Since there are $27$
unique time delays associated with each of the $\{a_i\}$ coefficients,
and because the gravitational wave signal enters in each TDI variable
at $6$ different times (the so called ``six-pulse response''), it
follows that in general the ZSS will display a total of ``$3 \times (6
\times 27) = 486$ pulses''. At the correct source location of course
all these pulses add-up to zero!

The analytic expressions for the $27$ time delays entering into
$a_{3}$ are given in the table below. The remaining $54$ can be
obtained from them by making a permutation over the spacecraft
indices.

\begin{table}
\caption{The twenty seven time delays entering into the function
  $a_3$, which multiplies the Fourier transform of the measurement
  $\gamma$. The other fifty four time delays, entering into the
  functions $a_1$ and $a_2$, can be obtained by
  permuting, once and twice respectively, the 
spacecraft indices of the expressions below.}
\begin{ruledtabular}
\begin{tabular}{ccccccccccc}
$k$ & & $\Delta_3^{(k)}$ & & $k$ & & $\Delta_3^{(k)}$ & & $k$ & & $\Delta_3^{(k)}$  \\
\hline
1 & & $L_1+L_2+2l\mu_3$ & & 10 & & $2L_1+2L_2+l(\mu_1+\mu_2)$ & & 19  & & $L_3+2l\mu_2$ \\
2 & & $L_1+L_2+2L_3+2l\mu_3$ & & 11 & & $L_1+2L_2+2L_3+l(\mu_1+\mu_3)$ & & 20 & & $2L_1+L_2+L_3+l(\mu_1+\mu_3)$ \\
3 & & $l(\mu_1+\mu_2)$ & & 12 & & $2L_1+L_2+2L_3+l(\mu_2+\mu_3)$ & & 21 & & $L_1+L_3+l(\mu_2+\mu_3)$ \\
4 & & $2L_3+l(\mu_1+\mu_2)$ & & 13 & & $L_1+2L_2+l(\mu_1+\mu_2)$ & & 22 & & $L_1+L_2+2l\mu_2$ \\
5 & & $L_1+l(\mu_1+\mu_3)$ & & 14 & & $2L_1+L_2+l(\mu_2+\mu_3)$ & & 23 & & $L_1+L_2+2L_3+2l\mu_2$ \\
6 & & $L_1+2L_3+l(\mu_1+\mu_3)$ & & 15 & & $L_1+L_2+L_3+l(\mu_1+\mu_2)$ & & 24 & & $L_3+2l\mu_1$ \\
7 & & $L_2+l(\mu_2+\mu_3)$ & & 16 & & $2L_1+2L_2+L_3+2l\mu_2$ & & 25 & & $2L_1+2L_2+L_3+2l\mu_1$ \\
8 & & $L_2+2L_3+l(\mu_2+\mu_3)$ & & 17 & & $L_1+2L_2+L_3+l(\mu_2+\mu_3)$ & & 26 & & $L_1+L_2+2L_3+2l\mu_1$ \\
9 & & $2L_1+2L_2+2L_3+l(\mu_1+\mu_2)$ & & 18 & & $L_2+L_3+l(\mu_1+\mu_3)$ & & 27 & & $L_1+L_2+2l\mu_1$ \\
\end{tabular}
\end{ruledtabular}
\end{table}

Finally, the functions $A_3^{(k)}$, introduced in equation
(\ref{expand1}), can be written in the following form
\begin{equation}
   A_3^{(k)} = \left[
   \Upsilon_{12}^{(k)} (F_{1}^{+}F_{2}^{\times} - F_{1}^{\times}F_{2}^{+})
+  \kappa_{31}^{(k)}(F_{3}^{+}F_{1}^{\times} -
   F_{3}^{\times}F_{1}^{+})
+ \chi_{23}^{(k)}(F_{2}^{+}F_{3}^{\times} -
   F_{2}^{\times}F_{3}^{+}) \right] \ ,
   \label{expandedSum}
\end{equation}
where the functions $\Upsilon_{12}^{(k)}$, $\kappa_{31}^{(k)}$, 
$\chi_{23}^{(k)}$ are

\begin{equation}
    \Upsilon_{12}^{(1)} = - \Upsilon_{12}^{(2)} = \frac{w_{1} - w_{2}}{2(1 - 
                    w_{1}^{2})(1 - w_{2}^{2})} \ ,
    \label{12Coeff1}
\end{equation}

\begin{equation}
    \Upsilon_{12}^{(3)} = \Upsilon_{12}^{(6)} = \Upsilon_{12}^{(8)} = - \Upsilon_{12}^{(4)}
    = - \Upsilon_{12}^{(5)} = - \Upsilon_{12}^{(7)} = \frac{1}{4(1 - w_{1})(1
    + w_{2})} \ ,
    \label{12Coeff3}
\end{equation}

\begin{equation}
    \Upsilon_{12}^{(9)} = \Upsilon_{12}^{(13)} = \Upsilon_{12}^{(14)} = -
    \Upsilon_{12}^{(10)} = - \Upsilon_{12}^{(11)} = - \Upsilon_{12}^{(12)} =
    \frac{1}{4(1 + w_{1})(1 - w_{2})} \ ,
    \label{12Coeff9}
\end{equation}

\begin{eqnarray}
   \Upsilon_{12}^{(15)} & = & \Upsilon_{12}^{(16)} = \Upsilon_{12}^{(17)} =
   \Upsilon_{12}^{(18)} = \Upsilon_{12}^{(19)} = \Upsilon_{12}^{(20)} =
   \Upsilon_{12}^{(21)} \nonumber \\ & = & \Upsilon_{12}^{(22)} = \Upsilon_{12}^{(23)} =
   \Upsilon_{12}^{(24)} = \Upsilon_{12}^{(25)} = \Upsilon_{12}^{(26)} =
   \Upsilon_{12}^{(27)} = 0
   \label{UpsilonZero}
\end{eqnarray}

\begin{equation}
    \kappa_{31}^{(3)} = -\kappa_{31}^{(5)} = -\kappa_{31}^{(12)} = \frac{1}{4(1 - w_{1})(1 - w_{3})} \ ,
    \label{13Coeff3}
\end{equation}

\begin{equation}
    \kappa_{31}^{(4)} = -\kappa_{31}^{(6)} = -\kappa_{31}^{(14)} = -\frac{1}{4(1 - w_{1})(1 + w_{3})} \ ,
    \label{13Coeff4}
\end{equation}

\begin{equation}
    \kappa_{31}^{(7)} = \kappa_{31}^{(11)} = -\kappa_{31}^{(9)} = -\frac{1}{4(1 + 
    w_{1})(1 + w_{3})} \ ,
    \label{13Coeff7}
\end{equation}

\begin{equation}
    \kappa_{31}^{(8)} = \kappa_{31}^{(13)} = -\kappa_{31}^{(10)} = \frac{1}{4(1 + w_{1})(1 - w_{3})} \ ,
    \label{13Coeff8}
\end{equation}

\begin{equation}
    \kappa_{31}^{(15)} = -\frac{w_{1} w_{3}}{(1 - w_{1}^{2})(1 - w_{3}^{2})}
    \ ,
    \label{13Coeff15}
\end{equation}

\begin{equation}
    \kappa_{31}^{(16)} = -\kappa_{31}^{(17)} = -\kappa_{31}^{(18)} = \frac{w_{3}}{2(1
    + w_{1})(1 - w_{3}^{2})} \ ,
    \label{13Coeff16}
\end{equation}

\begin{equation}
    \kappa_{31}^{(19)} = -\kappa_{31}^{(20)} = -\kappa_{31}^{(21)} = -\frac{w_{3}}{2(1
    - w_{1})(1 - w_{3}^{2})} \ ,
    \label{13Coeff19}
\end{equation}

\begin{equation}
        \kappa_{31}^{(22)} = -\frac{w_{1}}{2(1 - w_{1}^{2})(1 + w_{3})} \ ,
    \label{13Coeff22}
\end{equation}

\begin{equation}
    \kappa_{31}^{(23)} = \frac{w_{1}}{2(1 - w_{1}^{2})(1 - w_{3})} \ .
    \label{13Coeff23}
\end{equation}

\begin{equation}
    \kappa_{31}^{(1)} = \kappa_{31}^{(2)} = \kappa_{31}^{(24)} = \kappa_{31}^{(25)} =
    \kappa_{31}^{(26)} = \kappa_{31}^{(27)} = 0 \ .
    \label{13Coeff23}
\end{equation}


\begin{equation}
    \chi_{23}^{(3)} = -\chi_{23}^{(7)} = - \chi_{23}^{(11)} = \frac{1}{4(1 + w_{2})(1 + w_{3})} \ ,
    \label{23Coeff3}
\end{equation}

\begin{equation}
    \chi_{23}^{(4)} = -\chi_{23}^{(8)} = -\chi_{23}^{(13)} = -\frac{1}{4(1 + w_{2})(1 - w_{3})} \ ,
    \label{23Coeff4}
\end{equation}

\begin{equation}
    \chi_{23}^{(5)} = \chi_{23}^{(12)} = -\chi_{23}^{(9)} = -\frac{1}{4(1 - 
    w_{2})(1 - w_{3})} \ ,
    \label{23Coeff5}
\end{equation}

\begin{equation}
    \chi_{23}^{(6)} = \chi_{23}^{(14)} = -\chi_{23}^{(10)} = \frac{1}{4(1 - w_{2})(1 + w_{3})} \ ,
    \label{23Coeff6}
\end{equation}

\begin{equation}
    \chi_{23}^{15} = \frac{-w_{2} w_{3}}{(1 - w_{2}^{2})(1 - w_{3}^{2})} \ ,
    \label{23Coeff15}
\end{equation}

\begin{equation}
    \chi_{23}^{(17)} = \chi_{23}^{(18)} = -\chi_{23}^{(24)} = -\frac{w_{3}}{2(1 + w_{2})(1 - w_{3}^{2})} \ ,
    \label{23Coeff17}
\end{equation}

\begin{equation}
    \chi_{23}^{20} = \chi_{23}^{21} = -\chi_{23}^{25} = \frac{w_{3}}{2(1 - w_{2})(1 - w_{3}^{2})} \ ,
    \label{23Coeff20}
\end{equation}

\begin{equation}
    \chi_{23}^{26} = -\frac{w_{2}}{2(1 - w_{2}^{2})(1 + w_{3})} \ ,
    \label{23Coeff26}
\end{equation}

\begin{equation}
    \chi_{23}^{27} = \frac{w_{2}}{2(1 - w_{2}^{2})(1 - w_{3})} \ .
    \label{23Coeff27}
\end{equation}

\begin{equation}
    \chi_{23}^{(1)} = \chi_{23}^{(2)} = \chi_{23}^{(16)} = \chi_{23}^{(19)} = \chi_{23}^{(22)} = 
    \chi_{23}^{(23)} = 0 \ .
    \label{23Coeff27}
\end{equation}

\end{document}